 \documentclass[final,1p,times]{elsarticle}
\usepackage{graphics}
\usepackage{amsmath}
\usepackage{amssymb}
\usepackage[breaklinks=true,colorlinks=true,linkcolor=blue,urlcolor=blue,citecolor=blue]{hyperref}
\usepackage{soul,color}
\usepackage{epstopdf}
\usepackage{balance}

\biboptions{sort&compress}

\begin{document}
\begin{frontmatter}

\title{\textbf{Abrikosov fluxonics in washboard nanolandscapes}}

\author[1,2]{Oleksandr V. Dobrovolskiy}
\ead{Dobrovolskiy@Physik.uni-frankfurt.de}
\ead[url]{www.ovd.org.ua}
\address[1]{Physikalisches Institut, Goethe University, 60438 Frankfurt am Main, Germany}
\address[2]{Physics Department, V. Karazin Kharkiv National University, 61077 Kharkiv, Ukraine}

\begin{abstract}
Abrikosov fluxonics, a domain of science and engineering at the interface of superconductivity research and nanotechnology, is concerned with the study of properties and dynamics of Abrikosov vortices in nanopatterned superconductors, with particular focus on their confinement, manipulation, and exploitation for emerging functionalities. Vortex pinning, guided vortex motion, and the ratchet effect are three main fluxonic ``tools'' which allow for the dynamical (pinned or moving), the directional (angle-dependent), and the orientational (current polarity-sensitive) control over the fluxons, respectively. Thanks to the periodicity of the vortex lattice, several groups of effects emerge when the vortices move in a periodic pinning landscape: Spatial commensurability of the location of vortices with the underlying pinning nanolandscape leads to a reduction of  the dc resistance and the microwave loss at the so-called matching fields. Temporal synchronization of the displacement of vortices with the number of pinning sites visited during one half ac cycle manifests itself as Shapiro steps in the current-voltage curves. Delocalization of vortices oscillating under the action of a high-frequency ac drive can be tuned by a superimposed dc bias. In this short review a choice of experimental results on the vortex dynamics in the presence of periodic pinning in Nb thin films is presented. The consideration is limited to one particular type of artificial pinning structures --- directly written nanolandscapes of the washboard type, which are fabricated by focused ion beam milling and focused electron beam induced deposition. The reported results are relevant for the development of fluxonic devices and the reduction of microwave losses in superconducting planar transmission lines.
\end{abstract}

\begin{keyword}
Abrikosov fluxonics \sep washboard potential \sep nanopatterning \sep pinning \sep guiding \sep ratchet \sep vortex dynamics \sep niobium films
\end{keyword}

\end{frontmatter}

\tableofcontents

\section{Introduction}

\subsection{Retrospective}
The vast majority of technologically important superconductors are superconductors of type II: Magnetic field $\mathbf{H}$ penetrates these as a flux-line array of Abrikosov vortices \cite{Shu37etf,Shu08ujp,Abr57etp}. The repulsive interaction between vortices makes them to arrange in most cases into a triangular lattice with parameter $a_\bigtriangleup = (2\Phi_0/H\sqrt{3})^{1/2}$ ($\Phi_0$: magnetic flux quantum), which in the presence of a transport current can be either pinned (locally anchored) or be on the move. This depends on the counterbalance between the pinning and the driving (Lorentz) force $\mathbf{F}_L$ exerted on the vortices by the transport current. At currents densities larger than the critical (depinning) one $j > j_c$, that is when the Lorentz force dominates the pinning force, the vortex ensemble moves and faster vortex motion corresponds to larger dissipation. The associated resistive response enhances heat generation and charging times in superconducting integrated circuits.

The importance of controllable vortex \emph{pinning} for reducing the dissipation has been outlined in early \cite{Cam72aip} and later \cite{Bla94rmp,Bra95rpp} reviews. Technologically, an enhancement of $j_c$ can be achieved by strate\-gically positioning vortex pinning sites. Early attempts to improve the current-carrying ability of the samples through introducing artificial pinning sites date back to the late 1960th and relate to the pioneering works of Niessen and Wiessenfeld \cite{Nie69jap}, Morrison and Rose \cite{Mor70prl}, Fiory \cite{Fio71prl,Fio78apl}, Martinoli \cite{Mar75ssc} and many others. In addition, quantum interference effects (Shapiro steps) in rf-excited current-voltage curves (CVCs) were observed \cite{Fio71prl} for unpatterned Al films with weak pinning. Films showing stronger vortex pinning did not display observable steps. The effect was explained \cite{Fio71prl} by the nearly coherent motion of the vortex lattice in the presence of weak random pinning. The absence of the effect in films with stronger random pinning was attributed to the lack of said coherence. More pronounced interference steps in the CVCs were observed in thickness modulated Al films \cite{Mar75ssc,Fio78apl}.

A crossover from the weakly-dissipative regime at low frequencies to the strongly-dissipative regime at high frequencies was reported by Gittleman and Rosenblum \cite{Git66prl} who measured the rf and microwave (mw) power absorbed by vortices in non-patterned  PbIn  and  NbTa  films. The reason for this crossover lies in the shaking movement of vortices which are confined to one potential well, provided the ac drive occurs at sufficiently high frequency. This is in contrast to the dc-driven vortex dynamics where the motion of vortices can be described as their motion in a tilted pinning potential, i.\,e, when they visit many potential wells. The characteristic crossover frequency from the low-frequency (quasi-static or adiabatic) regime to the high-frequency regime is called the depinning frequency.

As a consequence of the spatial periodicity of the vortex lattice, there has been clear evidence \cite{Fio78apl,Mos10boo} that pinning is most effective when it is induced not by random but correlated (usually artificial) disorder and the period of the vortex lattice geometrically matches the period of the underlying pinning nanolandscape. Computer simulations revealed \cite{Luq07prb} that in this case the vortex-vortex interaction is effectively cancelled so that the dynamics of the entire vortex ensemble can be regarded as that of a single average vortex in an average pinning potential. Therefore, along with the formalism based on the time-dependent Ginzburg-Landau equation \cite{Kra78prl,Ros96prb,Von11prl,Sil11sst,Fom12nlt} for the superconducting order parameter, single-vortex mechanistic models relying upon the Langevin equation \cite{Cof12boo} were elaborated \cite{Git66prl,Cof91prl,Pom08prb,Maw97prb,Maw99prb,Shk99etp,Shk02prb,Shk06prb,Shk08prb,Shk09pcs,Shk09prb,Shk11prb,Dob12php,Shk12pcs,Shk12inb,Shk13ltp,Shk13snm,Shk14pcm,Shk14phc,Shk14pcs} and applied for the theoretical treatment of the vortex dynamics. The vortex-vortex interaction has also been considered in molecular dynamics simulations \cite{Zhu03prb,Zhu04prl,Luq07prb}. In these mechanistic models the vortices are regarded as point-like rigid entities.

Artificially created linearly extended pinning sites are known to be very effective for the reduction of the dissipation by vortices in one  or several particular directions. Indeed, if the pinning potential in a superconductor is anisotropic, the direction of vortex motion can be deflected away from the direction of the Lorentz force. In this case the nonlinear vortex dynamics becomes two-dimensional so that $\mathbf{v}  \nparallel \mathbf{F}_L$. The non-collinearity between the vortex velocity $\mathbf{v}$ and $\mathbf{F}_L$ is evidently more drastic the weaker the background isotropic pinning is \cite{Shk06prb}, which can otherwise mask this effect \cite{Sor05phd}. The most important manifestation of the pinning anisotropy is known as \emph{guided vortex motion}, or the guiding effect, meaning that vortices tend to move along the pinning potential channels rather than to overcome the potential barriers. As a consequence of the guided vortex motion, an even-in-field reversal transverse resistivity component appears, unlike the ordinary Hall resistivity which is odd with regard to the field reversal. Though a guiding of vortices can be achieved with different sorts of pinning landscapes \cite{Wor06pcs,Sil06inb} it is more strongly enhanced and can be more easily treated theoretically when using pinning potentials of the \emph{washboard} type.

One more intriguing effect appears when the pinning potential is asymmetric. In this case the reflection symmetry of the pinning force is broken and thus, the critical currents measured under current reversal are not equal. As a result, when subjected to an ac drive of zero mean value a net rectified motion of vortices occurs. This is known as the rocking \emph{ratchet effect} \cite{Plo09tas} which has been used, i.\,e. for removing the vortices from superconductors \cite{Lee99nat} as well as for studying the physics of a number of nanoscale systems, both solid state and biological \cite{Plo09tas,Han09rmp}.

All together vortex pinning, guided vortex motion, and the ratchet effect are the three main ``tools'' which allow for the dynamical (pinned or moving), the directional (angle-dependent), and the orientational (current polarity-sensitive) control over the fluxons, respectively. The respective domain of science and engineering is termed \emph{Abrikosov fluxonics} which emerged at the interface of superconductivity and nanotechnology \cite{Mos10boo,Mos11boo}. Abrikosov fluxonics is concerned with the study of properties and dynamics of vortices in nanopatterned superconductors with particular focus on their confinement, manipulation, and exploitation for emerging functionalities.

\subsection{Systems with washboard pinning nanolandscapes}

The distinctive feature of a washboard pinning potential (WPP) is that it is periodic in one dimension and constant in the perpendicular one. In contrast to systems with complex two-dimensional pinning potentials, for superconductors with a WPP an extensive theoretical description of the vortex dynamics is available. In its most general form it has been provided by Shklovskij and collaborators \cite{Shk99etp,Shk02prb,Shk06prb,Shk07pcs,Shk08prb,Shk09pcs,Shk09pcs2,Shk09prb,Shk09ltp,Shk10ltp,Shk11prb,Dob12php,Shk12pcs,Shk12inb,Shk13ltp,Shk13snm,Shk14ltp,Shk14pcm,Shk14phc,Shk14pcs}. From the viewpoint of theoretical modeling, saw-tooth \cite{Shk99etp,Shk02prb,Shk06prb,Shk09prb,Shk07pcs,Shk14ltp} and harmonic \cite{Shk08prb,Shk09pcs,Shk09pcs2,Shk11prb,Dob12php,Shk12pcs,Shk12inb,Shk13ltp,Shk13snm,Shk14pcm,Shk14phc,Shk14pcs} potentials represent the simplest WPP forms. On the one hand, these allow one to explicitly calculate the dc magnetoresistivity and the absorbed ac power as functions of the driving parameters of the experiment. On the other hand, WPPs are found across numerous experimental systems ranging from naturally occurring pinning sites in high temperature superconductors (HTSCs) to artificially created linearly-extended pinning sites in superconductor thin films. In fact, it is the latter which determine the functionality of present-day fluxonic devices.
\begin{figure*}[t!]
    \centering
    \includegraphics[width=1\textwidth]{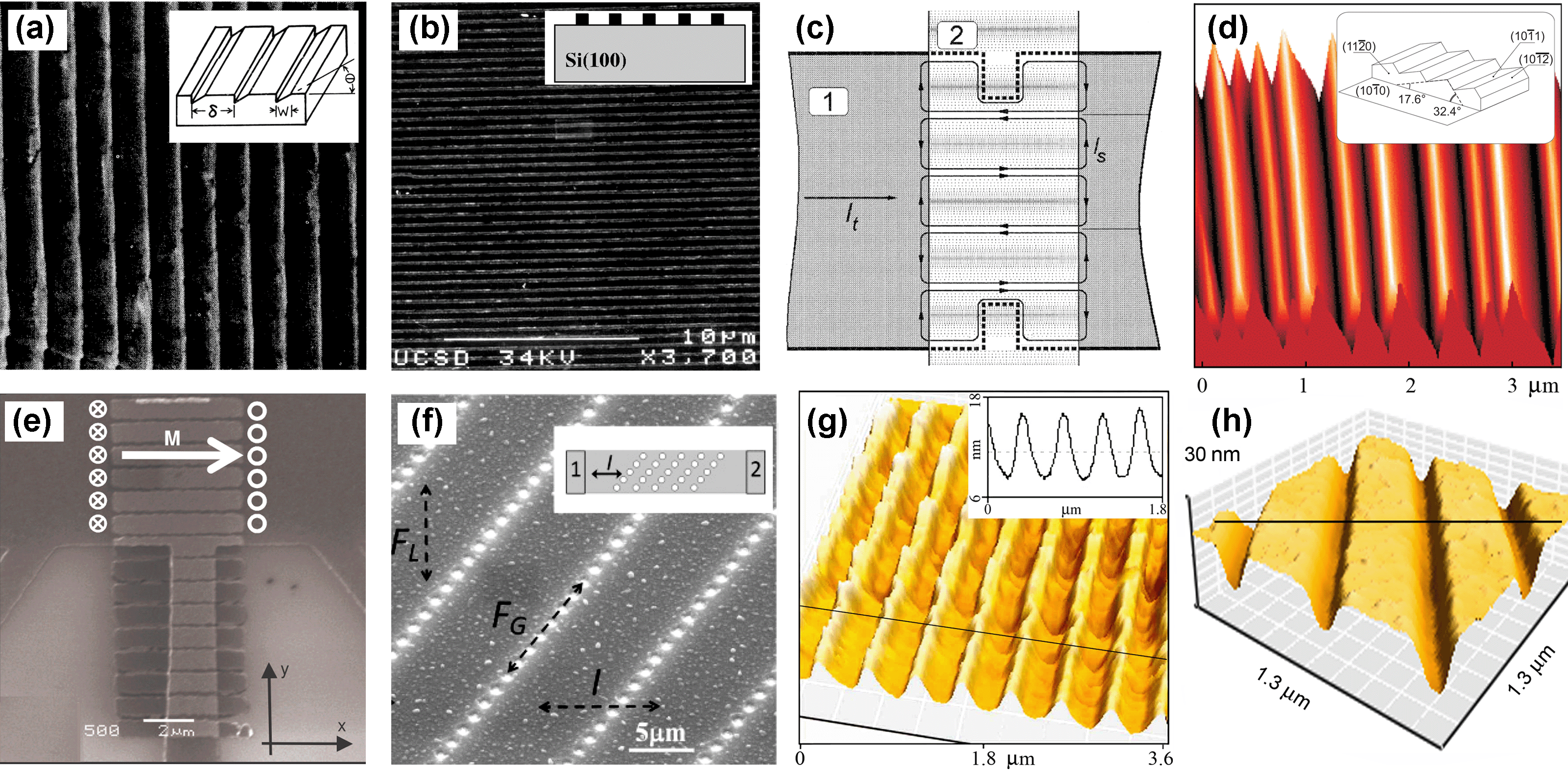}
    \caption{Exemplary systems with artificially created washboard pinning potentials: (a) In-$2\%$ Bi foil imprinted with diffraction grating. Adapted from \cite{Mor70prl}. (b) Parallel stripes of Ni prepared by electron-beam lithography on a Si substrate onto which a Nb film was sputtered. Adapted from \cite{Jaq02apl}. (c) Superconducting microbridge (1) with an overlaying magnetic tape (2) containing a pre-recorded magnetization distribution. Adapted from \cite{Yuz99pcs}. (d) Nb film deposited onto faceted $\alpha$-Al$_2$O$_3$ substrate surface. Adapted from \cite{Sor07prb}. (e) Al film deposited on top of $20$\,nm-thick Co bars. Adapted from \cite{Sil11sst}. (f) YBCO thin film with antidots milled by focused ion beam and arranged in a washboard-like fashion. Adapted from \cite{Wor12prb}. (g) Nb film surface with an array of Co nanostripes fabricated by focused electron beam-induced deposition \cite{Dob10sst}. (h) Nb film surface with an array of grooves etched by focused ion beam milling \cite{Dob12njp}. In addition, WPPs naturally occur in HTSCs either in form of intrinsic layers \cite{Ber97prl} or uniaxial twins \cite{Cha98sst,Pas99prl,Dan00prb}.}
    \label{fWPP}
\end{figure*}

A few representative systems with WPPs are exemplified in Fig. \ref{fWPP}. The first experimental realization of a WPP dates back to the work of Niessen and Weijsenfeld \cite{Nie69jap} who used a periodic modulation of the thickness of cold-rolled sheets of a Nb-Ta alloy. In their work the influence of isotropic pointlike disorder on the guiding of vortices was discussed for the first time. Morrison and Rose \cite{Mor70prl} used parallel microgrooves in the surfaces of homogeneous cast In-2\%Bi foils by pressing diffraction gratings into the foils. The pinning force thus induced was minimal for flux flow parallel to the grooves and maximal for perpendicular flux flow. Jacue \emph{et al} \cite{Jaq02apl} fabricated submicron Ni stripes by electron beam lithography underneath superconducting Nb films and showed that the stripes induce a strong anisotropy in the dissipative behavior. When the vortices moved perpendicular to the Ni stripes the resistance dropped by several orders of magnitude in comparison with the dissipation of vortices moving parallel to them. Yuzhelevski and Jung \cite{Yuz99pcs,Yuz99prb} applied a magnetic tape containing a pre-recorded magnetization distribution to the surface of high-$T_c$ thin films and thereby reversibly introduced controllable pinning profiles for the vortices therein. For instance, they demonstrated that periodic pinning conditions enforce a coherent flow of vortices that lead to the appearance of Shapiro steps in the CVCs.

Huth and collaborators used self-organization \cite{Hut02afm,Sor07prb} to provide semi-periodic, linearly extended pinning ``sites'' by spontaneous facetting of m-plane sapphire substrate surfaces on which Nb films were grown. They demonstrated that a pronounced guiding of vortices occurs. The experimental data \cite{Sor07prb} were in good agreement with the theoretical predictions within the model of competing isotropic and anisotropic pinning \cite{Shk06prb} that allowed the authors to fit their data to the theoretical expressions of Ref. \cite{Shk06prb}. Silhanek \emph{et al} \cite{Sil11sut} employed ferromagnetic bars densely packed in a linear array underneath a superconducting Al bridge to create two types of vortex chains of opposite polarity inside the superconductor. The authors observed, in particular, a drastic reduction of the dissipation in the channel populated with vortices having opposite polarity to the applied field and that the reduced dispersions in the velocity of vortices and their displacements in one individual row of magnetic bars leads to more pronounced Shapiro steps in the CVCs. W\"ordenweber \emph{et al} \cite{Wor12prb} used YBCO thin films with an array of blind holes (antidots) milled by focused ion beam and forming a quasi-washboard pinning landscape. They demonstrated that the mechanism of the guided flux transport depends on the microwave frequency and the geometrical size of the superconducting structures. Their crucial observation \cite{Wor12prb} is that vortex guiding persists up to high frequencies of several GHz.

The systems with naturally occurring pinning sites are largely represented by cuprates in which one distinguishes the intrinsic pinning induced by the layered structure itself \cite{Ber97prl} and the planar pinning caused by uniaxial twins \cite{Cha98sst,Pas99prl,Dan00prb}. Due to the high temperatures of the superconducting state and the short coherence lengths \cite{Bla94rmp}, natural pinning sites in cuprates have been shown to be less effective \cite{Sor05phd} as compared to regularly arranged artificial pinning structures.

\subsection{Processing of surfaces by focused particle beams}

An alternative approach to conventional lithographic techniques for the fabrication of pinning nanolandscapes is provided by two complementary mask-less nanofabrication techniques, namely focused ion beam (FIB) milling \cite{Utk08vst} and focused electron beam induced deposition (FEBID) \cite{Hut12bjn}. In the context of washboard nanolandscapes an advantage of these techniques lies in the possibility to directly write quasi-3D washboard nanolandscapes where the left-hand and right-hand slopes of the nanostructure's individual elements (grooves or stripes) can have different steepness. This results in an asymmetry of the induced pinning potential and leads to a difference in the depinning currents measured under current reversal \cite{Plo09tas}. A further benefit of the use of focused particle beams lies in their high-resolution (FIB: down to 25\,nm laterally and 0.5\,nm vertically \cite{Dob11snm}, FEBID of Co: down to 35\,nm laterally and 0.5\,nm vertically \cite{Dob11snm}). This is why FIB and FEBID are unique tools for the fabrication of washboard ratchet landscapes which are hard to obtain by other approaches. Atomic force microscope (AFM) images of exemplary pinning landscapes fabricated by FIB milling and FEBID of Co are shown in Fig. \ref{fWPP}(g,h) and in the insets to Fig. \ref{fCVC}(a,b).

\begin{figure}[t!]
    \centering
    \includegraphics[width=1\linewidth]{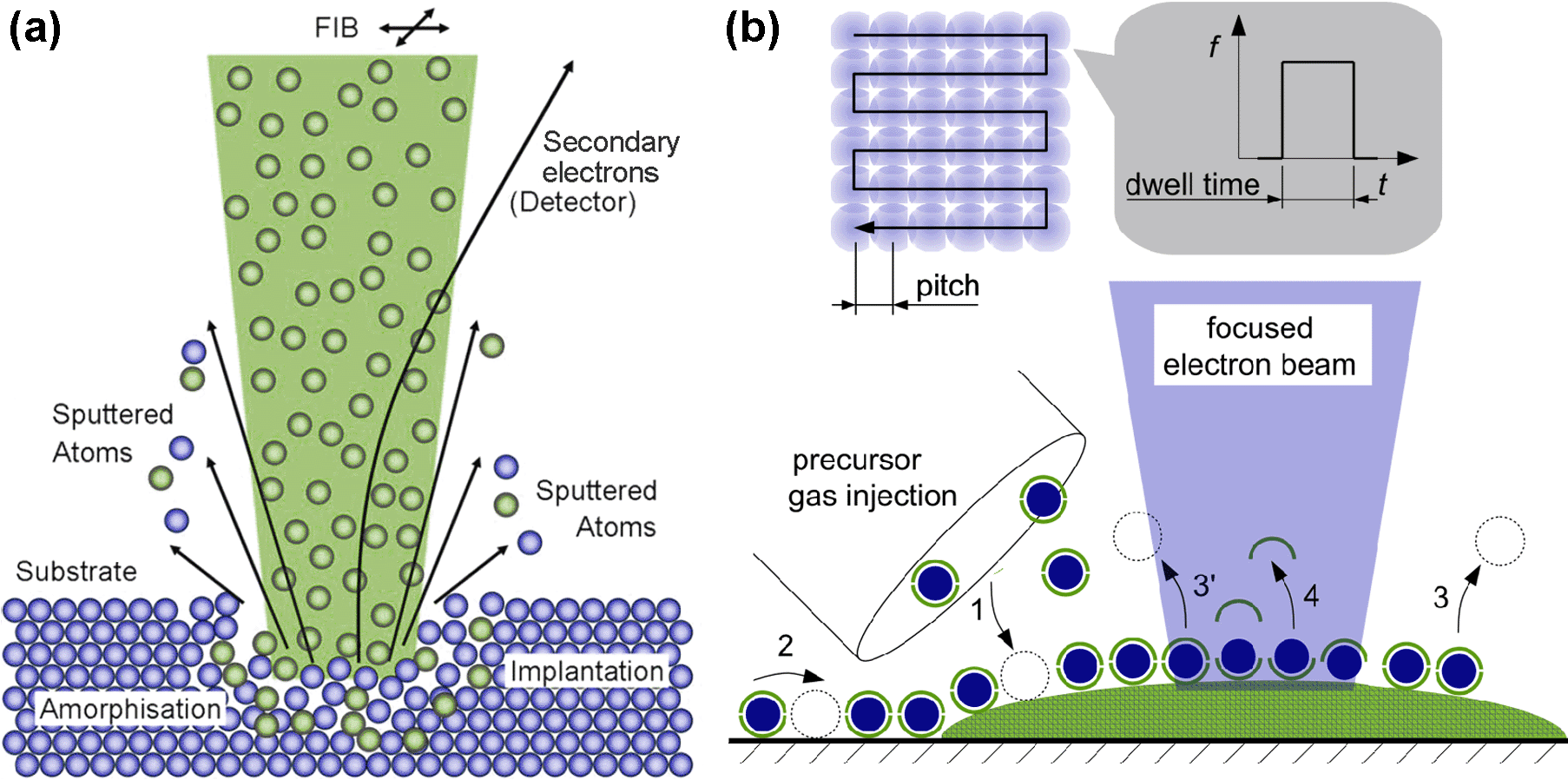}
    \caption{(a) Illustration of focused ion beam (FIB) milling. Removal of the material is accompanied by amorphization, implantation of Ga ions and vacancy generation in the processed surface. Adapted from \cite{Utk08vst} (b) Illustration of FEBID. Precursor molecules (here: organometallic complex; blue: metal, green: organic ligands) are supplied by a gas-injection system and physisorb (1) on the surface. Surface diffusion (2), thermally induced desorption (3) and electron-stimulated desorption (3') take place. Within the focus of the electron beam, adsorbed precursor molecules are (partly) dissociated followed by desorption of volatile organic ligands (4). Upper left: For pattern definition the electron beam is moved in a raster fashion (here: serpentine) over the surface and settles on each dwell point for a specified dwell time. After one raster sequence is completed the process is repeated until a predefined number of repeated loops is reached. Adapted from \cite{Hut12bjn}.}
    \label{fFEBIP}
\end{figure}

FIB milling and FEBID take place in the vacuum chamber of a scanning electron microscope (SEM), which in our case is a high-resolution dual-beam instrument Nova Nanolab 600 (FEI). FIB milling is done by Ga ions provided by a liquid metal ion source consisting of a tungsten needle covered with gallium. When heated to 30$^\circ$C, the liquid Ga reaches the needle tip and in the presence of a strong electric field forms a source cone with a diameter of several tens of nm \cite{Nak93jpd}. The most important processes which take place upon interaction of ions with matter are nonelastic scattering with electrons and elastic scattering with nuclei, which leads to the braking of ions and their possible implantation. The principle of FIB milling is sketched in Fig. \ref{fFEBIP}(a). The ion penetration depth is determined by the ion energy ($1\div30$\,keV typically) and the atomic number of the constituent atoms. An early report on the use of FIB-milled trenches to increase the critical current in Nb films can be found in \cite{Pau04prb}.

FEBID is a process of decomposing adsorbed molecules using a focused beam of electrons to form a deposit on a substrate. Well-established applications of FEBID range from photomask repair \cite{Lia05vst} to fabrication of nanowires \cite{Kom14jap}, nanopores \cite{Dan06lan}, magnetic \cite{Gab10nan} and strain \cite{Sch10sen} sensors as well as direct-write superconductors \cite{Mak14nsr,Win14apl}. When the material to be deposited is magnetic, it is possible to tune the deposit's magnetic properties which is essential for basic research in micro-magnetism \cite{Lar14apl} and spin-dependent transport \cite{Kei06nat} as well as for such applications as magnetic domain-wall  logic \cite{All05sci} and memory \cite{Par08sci}, fabrication of Hall sensors \cite{Gab10nan} and cantilever tips \cite{Bel12rsi} for magnetic force microscopy (MFM). In particular, the ability to tune the magnetization is the basic property needed for the realization of stacked nanomagnets \cite{Tak06jap}, pinning of magnetic domain walls \cite{Bri11prl} and Abrikosov vortices \cite{Vel08mmm}, magnetic sensing \cite{Gab10nan} and storage \cite{Par08sci}, and the necessary condition for spin-triplet proximity-induced superconductivity \cite{Buz05rmp,Ber05rmp,Wan10nat,Kom14apl}. Early reports on the use of FEBID for the fabrication of pinning nanostructures can be found in \cite{Dob10sst,Dob11pcs}.

The principle of FEBID is schematically shown in Fig. \ref{fFEBIP}(b). The precursor gas [in this case Co$_2$(CO)$_8$] is supplied to the SEM chamber in the vicinity of the substrate surface by a gas injection system. The precursor molecules adsorb, desorb and diffuse onto the substrate surface. Given optimal conditions applied during the deposition process, the surface is covered homogeneously and permanently with precursor gas molecules. Further, by rastering over the substrate surface with the electron beam, the adsorbed precursor molecules are excited and, as a result, dissociated. During the dissociation process, decomposition of the precursor molecules to volatile and non-volatile components takes place. The volatile components are pumped away from the SEM chamber, while the non-volatile components are deposited. Scanning takes place over a predefined area, resulting in deposits that can be freely configured in the desired spatial geometry \cite{Pla15prc}. Relevant FEBID parameters are the primary beam energy ($1\div30$\,keV), the beam current ($10$\,pA $\div10$\,nA), the distance between successive dwell points of the electron beam (pitch, $10\div100$\,nm), and the time period over which the electron beam is helt at each dwell point (dwell time $100$\,ns$\div10$\,ms). The variation and combination of the given FEBID parameters have a significant impact on precursor decomposition efficiency, deposition rate, and composition of the resulting deposits. The metal content of as-deposited FEBID structures is strongly dependent on the applied precursor gas and the deposition parameters. A long standing problem with the use of organo-metallic precursors is the low efficiency of the process, which results in an abundance of carbon with inclusions of oxygen and a low metal content. The problem, however, can be partially solved by post-growth annealing and electron irradiation of the deposits \cite{Beg15nan,Dob15bjn}.

To summarize the introductory part, the bloom of studies of the dynamics of vortices in nanopatterned superconductors \cite{Mos10boo,Mos11boo} has led to the appearance of a new field of science and engineering termed Abrikosov fluxonics. In this, washboard nanolandscapes represent a unique class of pinning structures capturing the three main fluxonic effects, namely vortex pinning, guiding and the ratchet effect which can be treated in the framework of mechanistic models. On the experimental side, pre-defined symmetric and asymmetric WPPs can be fabricated by processing superconductor surfaces by focused particle beams. A choice of effects generic to the vortex dynamics in WPPs is presented next.

\section{Vortex dynamics in washboard nanolandscapes}

\subsection{Resistance anisotropy in Nb thin films with Co nanostripes}

\begin{figure}[t!]
    \centering
    \includegraphics[width=0.7\linewidth]{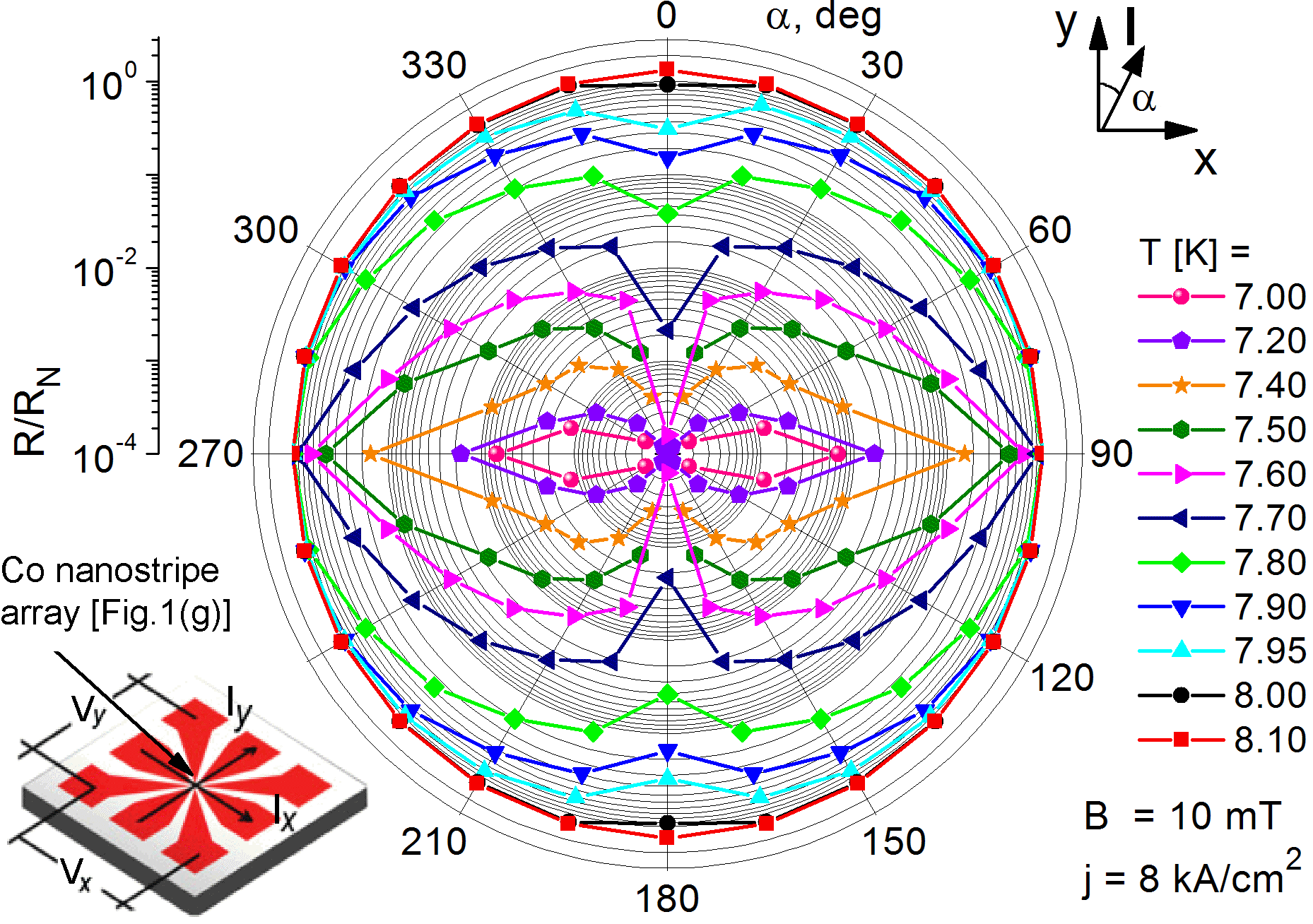}
    \caption{Polar diagram of the total magnetoresistance of a Nb films with a Co nanostripe array upon rotating the total current. $R_N$ designates the total resistance of the film in the normal state. The experimantal geometry is shown in the insets. After \cite{Dob10sst}.}
    \label{fRvAlpha}
\end{figure}

Employing the current-oriented setup used by Pastoriza \emph{et al} \cite{Pas99prl} to the thin-film geometry it was shown that a Co nanostripe array [Fig. \ref{fWPP}(g)] deposited on top of Nb films induces a strong WPP \cite{Dob10sst,Dob11pcs,Dob12ppa}. The film contained eight contacts as shown in the left bottom part of Fig. \ref{fRvAlpha}. The two current components $I_x$ and $I_y$ applied along the $x$- and $y$-axis allowed one to rotate the total current vector  $\mathbf{I}$ with respect to the guiding direction of WPP, which is parallel to the $y$-axis. Two voltage components $V_x$ and $V_y$ were measured simultaneously. This allowed one to plot the total resistance $R= (V_x^2 + V_y^2)^{1/2}/(I_x^2 + I_y^2)^{1/2}$ as a function of the angle $\alpha$ between $\mathbf{I}$ and the $y$-axis. Though a quantitative analysis of the vortex response in such a geometry is known to be complicated due to the complex current distribution in the cross-strip constriction \cite{Sil08apl}, this scheme suited well to verify whether the Co-FEBID stripes invoke any anisotropy in the resistive response in principle. The period of the Co nanostripe array was $400$\,nm, the stripe height was $8\pm0.5$\,nm, and the nanopattern was fabricated within a cross area of $46\times46\,\mu$m$^2$.

Polar diagram of the total resistance is shown in Fig. \ref{fRvAlpha} for a series of temperatures from 7.00 to 8.10\,K ($0.86\div0.99T_c$) and the magnetic field $B = 10$\,mT resulting in one row of vortices pinned at the Co stripes. At lower temperatures the resistive response is substantially anisotropic, whereby the minimal value of the resistance is attained at $\alpha = 0^\circ$ and $180^\circ$. The difference between the resistance maximum and minimum values is up to four orders of magnitude. When approaching the superconducting transition temperature $T_c$ the response becomes isotropic since a diffusion-like mode strongly dominates in the vortex motion. Noting that the Lorentz force is oriented perpendicular to the transport current direction, the 90$^\circ$ current orientation corresponds to the vortices moving along the Co stripes, whereas the 0$^{\circ}$ orientation corresponds to their transverse motion. The observed maximum of the resistance for 90$^\circ$ and the minimum for 0$^{\circ}$ prove that at low temperatures the vortices prefer to move along the pinning channels induced by the stripes of Co. As far as a faster vortex motion corresponds to a larger resistance, such an eight-shaped form is a clear signature of a guided vortex motion along the Co stripes. Hence, cross-strips with a WPP can be used as fluxonic valves to govern the resistive response in superconducting planar transmission lines.

\subsection{Adjustable guiding of vortices by magnetic field tuning}
Another approach to study the vortex guiding effect consists not in rotating the current direction with respect to the pinning channels but in tilting the pinning structure with respect to the current direction in the conventional four-probe geometry \cite{Dob12njp,Dob13pcs}. Thus, by means of a single deposition process two nominally identical epitaxial Nb (110) films with a thickness of $52$\,nm were prepared by dc magnetron sputtering \cite{Dob12tsf} onto ($11\bar{2}0$) sapphire substrates cut from one and the same $\alpha$-Al$_2$O$_3$ wafer. Each film contained four $30\times100\,\mu$m$^2$ bridges so that both films accomodated seven replica of the $450$\,nm-periodic FIB-milled profile shown in Fig. \ref{fWPP}(h) and tilted at an angle $\alpha$ of $0^{\circ}, 15^{\circ}, 30^{\circ}, 45^{\circ}, 60^{\circ}, 75^{\circ}$, $90^{\circ}$ with respect to the current direction as well as one bridge left non-patterned.

Resistance dips and critical current maxima were observed at the magnetic fields $H = 8.8$\,mT and $11.7$\,mT \cite{Dob12njp}. The geometrical relation between the assumed triangular vortex lattice with parameter $a_\bigtriangleup = (2\Phi_0/H\sqrt{3})^{1/2}$ and the matching conditions $a_\bigtriangleup = 2a/\sqrt{3}$ and $a_\bigtriangleup = a$ allowed us to study adjustable guiding of vortices by fine-tuning magnetic field. To exclude the Hall-like contribution in the longitudinal resistivity component \cite{Shk06prb}, the even-in-field resistivity was measured. Three qualitatively different cases were studied: First, the field of $H_a = 8.8$\,mT corresponded to the fundamental matching field, i.\,e., when all vortices were pinned at the groove bottoms and there were no interstitial vortices. Second, a field of $H_b = 11.7$\,mT represented the partial matching field at which half the total number of vortices were pinned at the groove bottoms and the remaining vortices were pinned by randomly distributed isotropic pins in between. Finally, a field of $H_c = 15$\,mT was used as a mismatching field for reference.
\begin{figure}[t!]
    \centering
    \includegraphics[width=0.5\linewidth]{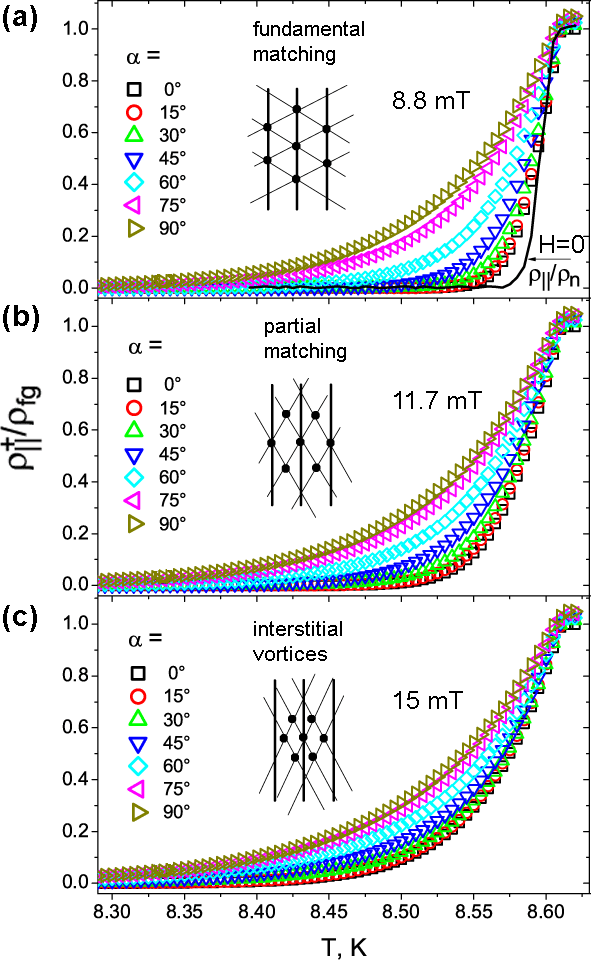}
    \caption{The temperature dependence $\rho^+_{\parallel}(T)$ for different grooves tilt angles $\alpha$ with respect to the transport current at three magnetic field values. After \cite{Dob12njp}.}
    \label{fGuiding}
\end{figure}

The temperature dependences of the resistivity components thus measured are shown in Fig. \ref{fGuiding}. The arrangements of vortices with respect to the underlying pinning nanolandscapes are shown in the respective insets. The \emph{tunability} of the guiding effect is clearly seen, namely the guiding effect becomes less pronounced upon detuning the field value away from the fundamental matching configuration. In doing so, the most likely pinning sites along which the flux lines move through the samples could be selected. By this, either the background isotropic pinning of the pristine film or the enhanced isotropic pinning originating from the nanoprocessing was probed.

An Arrhenius analysis of the resistivity data allowed us to quantify the pinning activation energy $U\simeq6000$\,K at the groove bottom for the transverse vortex motion, while $U$ was about $1000$\,K for the longitudinal motion. The activation energy for the isotropic pinning in the processed films was estimated as $U\simeq830$\,K between the grooves, while $U$ amounted to $\simeq760$\,K in the non-patterned sample. The pinning activation energies were found to correlate well with the results of local compositional analysis by energy-dispersive  x-ray spectroscopy, that allowed us to elucidate the pinning mechanisms at work in superconductors with FIB-milled nanolandscapes \cite{Dob12njp}.

\subsection{Synchronization effects under combined dc an ac drives}

\begin{figure}[t!]
    \centering
    \includegraphics[width=0.7\linewidth]{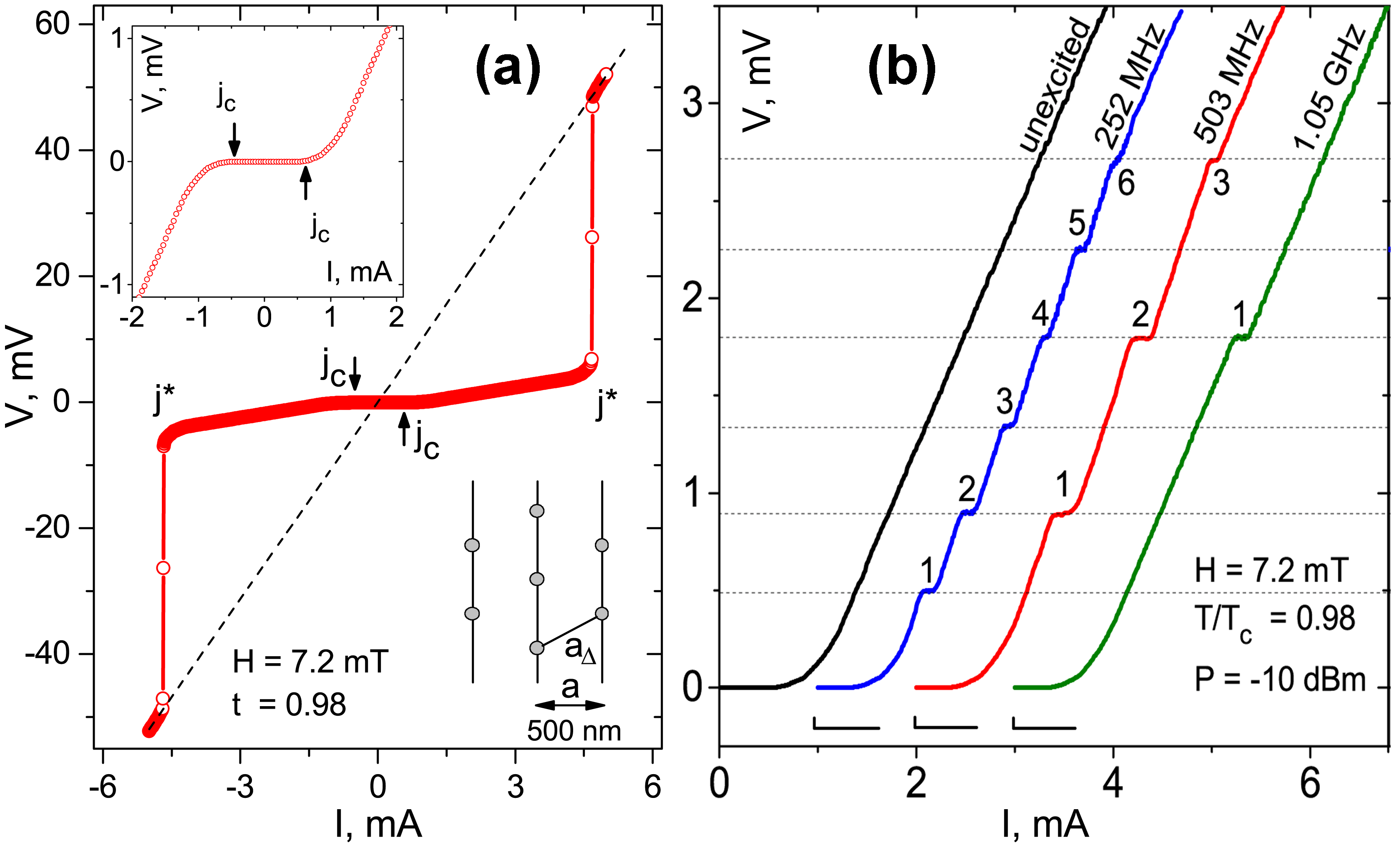}
    \caption{(a) CVCs of the microstrip at the fundamental matching field $H = 7.2$\,mT and the reduced temperature $t=0.98$. The dashed line is a guide for the eye. Upper inset: The CVC in the small currents regime. The arrows indicate the depinning currents determined by the $0.1\,\mu$V voltage criterion. Lower inset: The arrangement of vortices at $7.2$\,mT corresponding to the fundamental matching configuration, where $a_\bigtriangleup$ is the vortex lattice parameter and $a$ is the nanopattern period. (b) Manifestation of the Shapiro steps in the flux-flow regime of the CVCs for a series of ac frequencies, as indicated. The origins of the curves are shifted by $1$\,mA to the right along the current axis. After \cite{Dob15snm}.}
    \label{fShapiro}
\end{figure}

The vortex dynamics under combined dc and high-frequency ac drives was investigated  in nanopatterned Nb microstrips by broadband transmission spectroscopy using a custom-made sample probe \cite{Dob15mst}. The unexcited CVC of a Nb microstrip with a $500$\,nm-periodic WPP is shown in Fig. \ref{fShapiro}(a). The depinning current densities are nearly equal to $6.4$\,kA/cm$^2$ for both CVC branches. This is due to the cross-sectional symmetry of the symmetric pinning nanolandscape in Fig. \ref{fCVC}(a) which will be referred to as sample S. At $j > j_c$ one recognizes the linear regime of viscous flux flow. At yet larger current densities, $j^\ast \approx 65$\,kA/cm$^2$, abrupt jumps to the normal state follow. The fact that the CVCs does not exhibit a continuous crossover from the flux-flow resistance to the normal-state resistance is associated with non-equilibrium phenomena in the vortex motion with $j^\ast$ being the Larkin-Ovchinnikov instability current density \cite{Lar75etp}. In the following considered are the CVCs at $j<j^\ast$.

The modification of the CVCs in the presence of a mw excitation \cite{Dob15snm} is shown in Fig. \ref{fShapiro}(b). The CVCs exhibit Shapiro steps which appear due to the synchronization of the motion of Abrikosov vortices to the microwave frequency. The steps occur at voltages \cite{Fio71prl} $V = n V_0 \equiv nN\Phi_0 f$, where $n$ is an integer, $N$ is the number of vortex rows between the voltage leads, $f$ is the microwave frequency, and $\Phi_0 = 2.07\times10^{-15}$\,Vs is the magnetic flux quantum. The steps in the CVCs arise when one or a multiple of the hopping period of Abrikosov vortices coincides with the period of the ac drive.

In Fig. \ref{fShapiro}(b) one can distinguish up to six lowest-order Shapiro steps (refer to the curve for $252$~MHz) while higher-order steps are smeared. Given the geometrical dimensions of the microstrip and the fundamental matching field configuration for a triangular vortex lattice [see the inset to Fig. \ref{fShapiro}(a), the expected number of vortex rows between the voltage leads is equal to $N=866$. The fact that the experimental data could be fitted \cite{Dob15snm} with the same $N$ suggests that all vortices move coherently. This strongly coherent motion is caused by both, the high periodicity of the nanogroove array and a relatively weak contribution of the background isotropic pinning due to structural imperfectness as compared to the dominating strong pinning owing to the nanopatterning. When tuning the field value away from the matching configuration the steps disappear.

\subsection{Vortex lattice matching effects in absorbed ac power}
\begin{figure}[b!]
    \centering
    \includegraphics[width=0.68\linewidth]{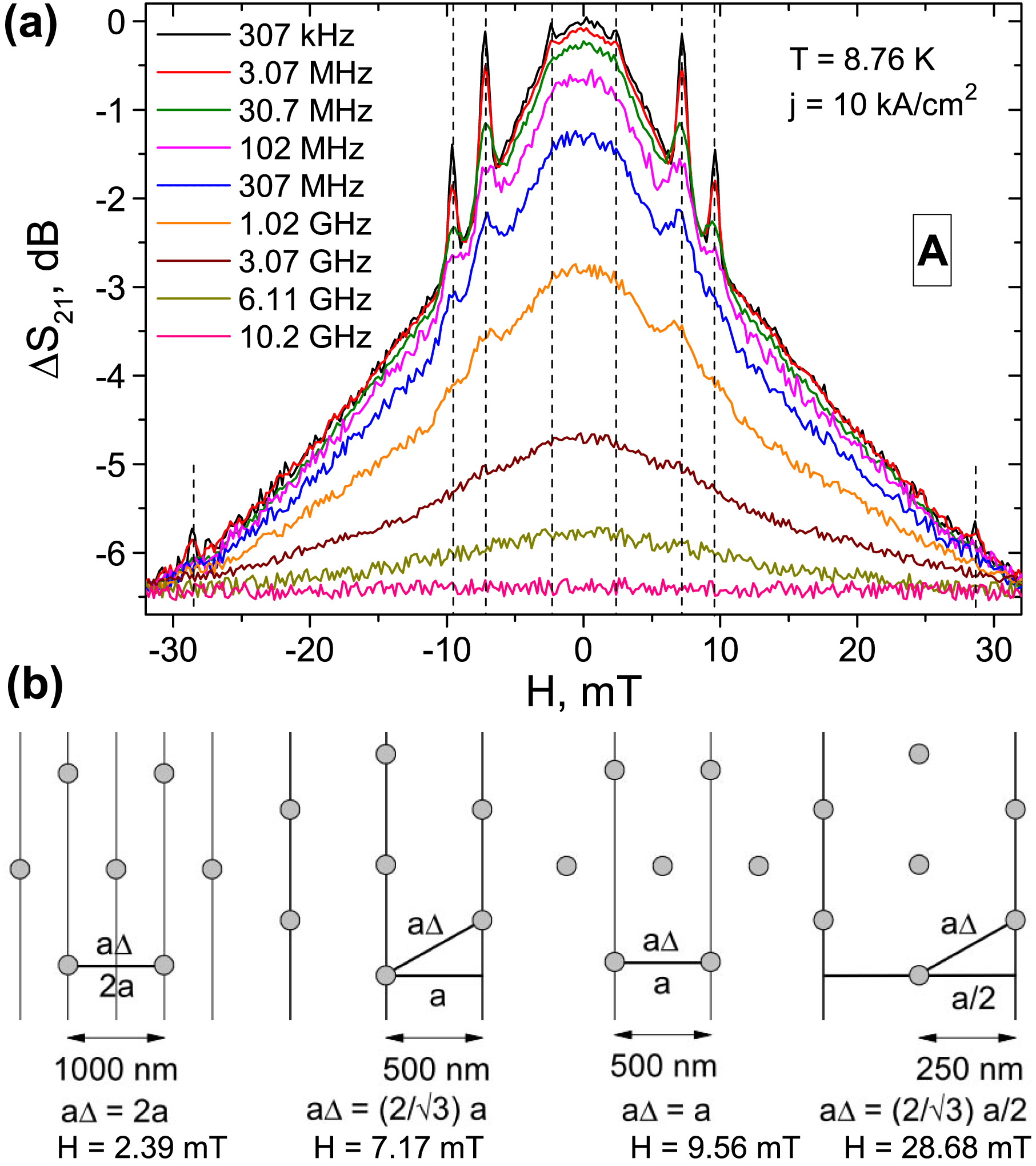}
    \caption{(a) Changes in the forward transmission coefficient $\Delta S_{21}$ of sample A under magnetic field reversal. (b) Vortex lattice configurations corresponding to the observed peaks in panel (a) along with the matching conditions $a_\bigtriangleup = ka$ and the calculated magnetic fields values. After \cite{Dob15apl}.}
    \label{fMatching}
\end{figure}

Figure \ref{fMatching}(a) presents the relative change of the forward transmission coefficient $\Delta S_{21} $ for a Nb microstrip with asymmetric grooves as a function of the magnetic field at $T = 0.98T_c$ and a series of frequencies. Here, $\Delta S_{21}$ is a measure for the mw loss due to vortex motion. An AFM image of the WPP landscape is shown in Fig. \ref{fCVC}(b); this microstrip will be referred to as sample A. The $\Delta S_{21}(H)$ data attest to that at lower frequencies, the mw loss rises with increasing magnetic field, whereas $\Delta S_{21}$ becomes less sensitive to the field variation at elevated frequencies and saturates at the $-6.4$\,dB level (maximal loss). A detailed inspection of $\Delta S_{21}(H)$ unveils a pronounced reduction of the mw loss (peaks in $\Delta S_{21}$) at $7.2$\,mT and $9.6$\,mT, while less pronounced peaks (recognizable at $f\lesssim 3$\,MHz) correspond to $2.4$\,mT and $28.7$~mT. An increase of the microwave power by $20$\,dB leads to the disappearance of these matching field effects as the sample is driven to the normal state.

Commensurability effects in the mw power were recently observed also in Pb films with square pinning site arrays by magnetic field-dependent mw power reflection spectroscopy~\cite{Sol14prb} and broadband permeability transmission measurements \cite{Lar15nsr}. In the present case, for the assumed triangular vortex lattice with lattice parameter $a_\bigtriangleup = (2\Phi_0/H\sqrt{3})^{1/2}$ and the geometrical matching conditions $a_\bigtriangleup = ka$, the arrangement of vortices with respect to the underlying pinning nanolandscape is shown in Fig.~\ref{fMatching}(b). The calculated field values for the triangular vortex lattices agree very well with those deduced from Fig.~\ref{fMatching}(a), although one cannot exclude the presence of more complex arrangements at fields 9.6\,mT and 28.7\,mT as suggested by computer simulations \cite{Luq07prb}. The field $7.2$\,mT corresponds to the maximal vortex density when each vortex is pinned at the bottom of a groove and there are no interstitial vortices, i.\,e. to the \emph{fundamental} matching configuration. It is this field for which the following effects are reported.

\subsection{Microwave filter exploiting vortex ratchet effect}
\begin{figure}[b!]
\centering
    \includegraphics[width=0.7\linewidth]{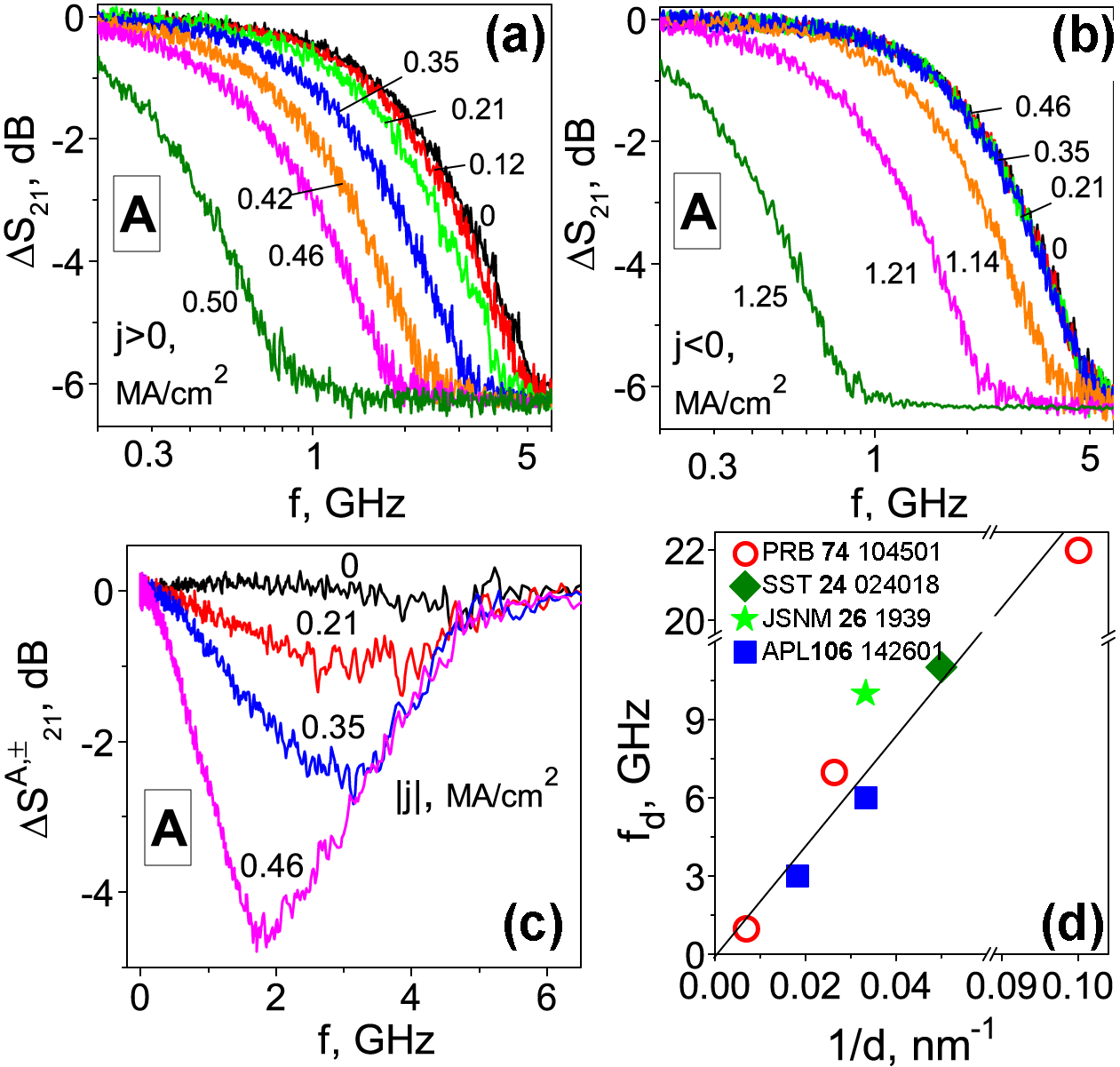}
    \caption{Dependence $\Delta S_{21}(f)$ of sample A at positive (a) and negative (b) dc densities at $H = 7.2$\,mT, $T = 0.3T_c$, and $P = - 20$\,dBm. (c) Difference signal $\Delta S^{A,\pm}_{21}(f) = \Delta S_{21}(j) - \Delta S_{21}(-j)$. (d) Zero-bias depinning frequency at $H=0$ and $T=0$ as a function of the (inverse) Nb film thickness for the data of Refs. \cite{Jan06prb,Sil11sst,Pom13snm,Dob15apl}. The straight line $f_d \propto 1/d$ is guide for the eye. After \cite{Dob15apl}.}
    \label{power}
\end{figure}

Figure~\ref{power}(a) and (b) show the frequency dependence $\Delta S_{21}(f)$ of sample A for the same magnitudes of positive and negative dc current densities. In the absence of a dc bias, the mw loss is maximal at high frequencies, whereas the vortex response is weakly dissipative at low frequencies. For both dc bias polarities, upon increasing the dc value, the $\Delta S_{21}$ curves shift towards lower frequencies but the magnitudes of the shifts substantially differ, as is in contrast with sample S \cite{Dob15apl}.

\begin{figure}[b!]
    \centering
    \includegraphics[width=0.9\linewidth]{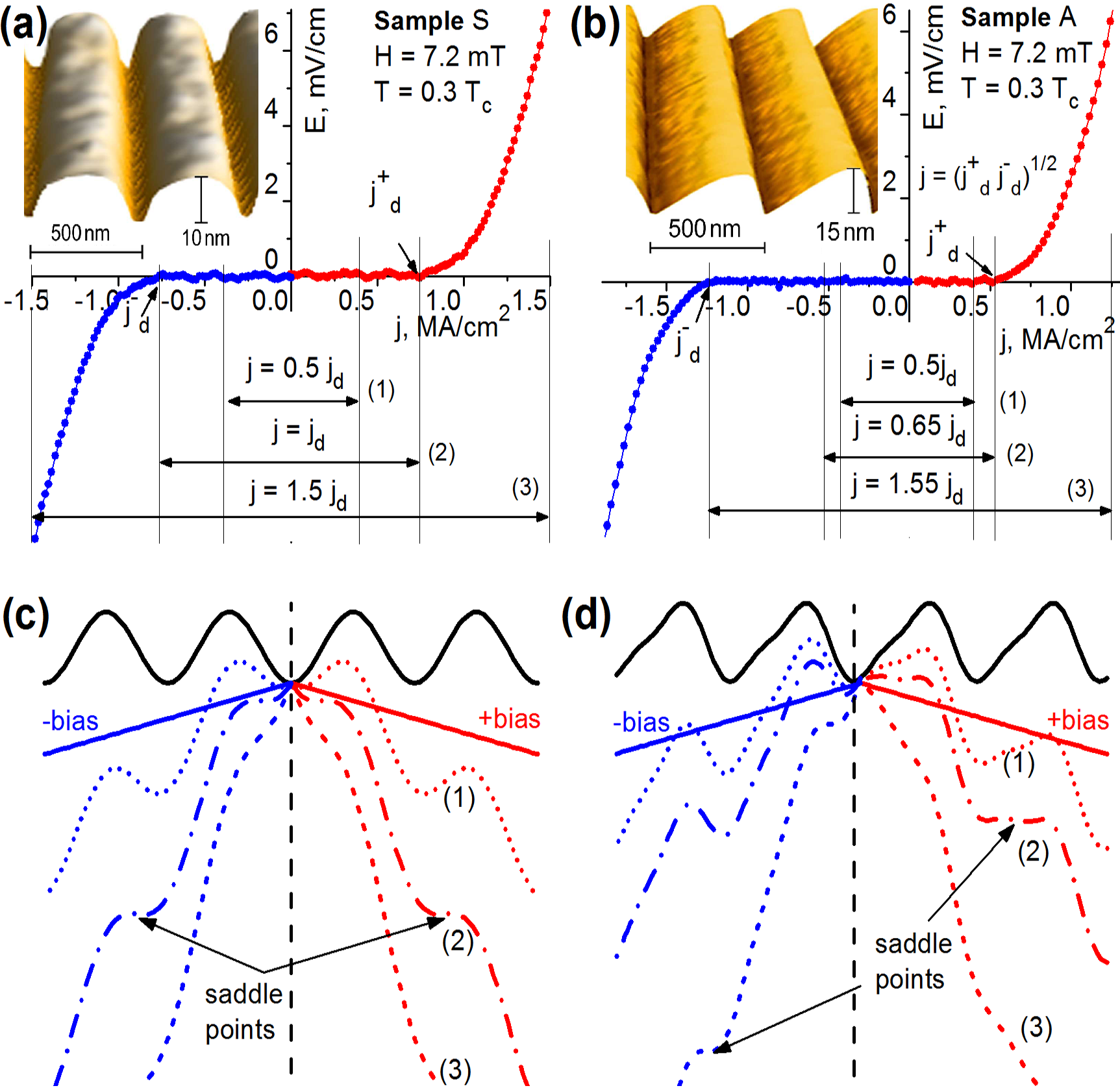}
    \caption{CVCs for samples S (a) and A (b) at $T = 0.3T_c$ and $H = 7.2$\,mT, respectively. The qualitatively different regimes in the vortex dynamics ensue upon application of currents depicted by horizontal arrows. Tilts of the washboard potentials (c) and (d) sketched for the ac amplitudes (1)--(3) in the CVCs (a) and (b), respectively. The regimes (2) for both samples are used for the presentation of tailoring the excess loss level in Fig. \ref{fModul}. After \cite{Dob15sbm}.}
    \label{fCVC}
\end{figure}

The reduction of the depinning frequency upon increasing the dc bias can be understood as a consequence of the effective lowering of the pinning potential well due to its tilt by the dc current, see also Fig. \ref{fCVC}(c). 
Indeed, the mechanistic consideration \cite{Shk11prb} of a vortex as a particle leads to the conclusion that during an ac semiperiod, while the pinning potential well is broadening, with increasing $f$ the vortex has no longer time $(\sim 1/f)$ to reach the areas where the pinning forces dominate and, hence, the response becomes stronger dissipative already at lower frequencies, as compared with the zero-bias curve. The same mechanistic scenario can be applied for the explanation of the difference in the shifts of the depinning frequencies for the positive and the negative dc biases, caused by the different groove slope steepnesses \cite{Shk14pcm}, refer to Fig. \ref{fCVC}(d). Thus, by electrical resistance measurements it was revealed that the groove asymmetry causes a difference in the depinning currents ($j_d$) for the positive and the negative branches of the CVC for sample A, see Fig. \ref{fCVC}(b). At $T=0.3T_c$ and $H=7.2$\,mT these amount to $0.52$\,MA/cm$^2$ and $1.25$\,MA/cm$^2$, respectively. This is in contrast to the CVC of sample S which is symmetric with $j_d = 0.75$\,MA/cm$^2$, see Fig. \ref{fCVC}(a). Here, $j_d$ is determined by the $10\,\mu$V/cm electric field strength criterion. Hence, sample A exhibits a microwave filter behavior, whose cut-off frequency depends not only on the dc value but also on the dc polarity.

The difference signal $\Delta S^{A,\pm}_{21}(f) = \Delta S_{21}(j) - \Delta S_{21}(-j)$ for sample A is plotted in Fig.~\ref{power}(c). One clearly sees the difference between the mw loss measured for the same, positive and negative dc bias magnitudes. This is caused by the fact that at moderate dc values (with respect to the corresponding $j_d$), for the gentle-slope direction of the vortex motion the depinning frequency is markedly lower than for the steep-slope direction. Specifically, the effect is most pronounced at $j = 0.46$\,MA/cm$^2$ at which the depinning frequencies are $f_d\approx1$\,GHz and $f_d\approx2.7\simeq3.02$\,GHz $= f_d(j=0)$ for the gentle-slope and the steep-slope direction, respectively. The depinning frequencies are determined at the $-3$\,dB excess loss level.

The frequency characteristics of the filter $\Delta S_{21}(f)$ can be fitted to the expression $\Delta S_{21}(f) = 1/[1 + (f_{d}/f)^n]$ with the exponent $n=2$ for $0 < j <0.42$\,MA/cm$^2$ and $n\approx1.85$ for $j\simeq j_d$ for sample S. For the gentle-slope direction of sample A $n\approx2.1$ and for its steep-slope direction $n\approx1.9$. In general~\cite{Poz11boo}, $n=2$ corresponds to a first-order filter roll-off $-10\log_{10}[(f/f_d)^2]$. The mw filter operates under the fundamental matching field condition and the observed effects become entangled and eventually vanish upon tuning the field value away from 7.2\,mT.

In Fig. \ref{power}(d) is plotted the zero-bias depinning frequencies at $T=0$ and $H = 0$, $f_d(0,0)$, for both samples \cite{Dob15apl} along with the data for Nb films reported in Refs. \cite{Jan06prb,Sil11sst,Pom13snm}. The values $f_d(0,0)$ were estimated by the expression $f_d(T) = f_d(0)[1-(T/T_c)^4]$ successfully used~\cite{Zai03prb} for fitting the experimental data in high-$T_c$ films and the empiric dependence $f_d(H) = f_d(0)[1-(H/H_{c2})^2]$ observed~\cite{Jan06prb} for Nb films, respectively. For samples S and A, their thicknesses were taken after subtracting the groove depths. From Fig.~\ref{power}(d) it follows that the cumulative data for Nb films can be fitted to the phenomenological law $f_d \propto 1/d$, where $d$ is the film thickness.

\subsection{Synthesis of arbitrary microwave loss levels}
\begin{figure}[b!]
    \centering
    \includegraphics[width=0.6\linewidth]{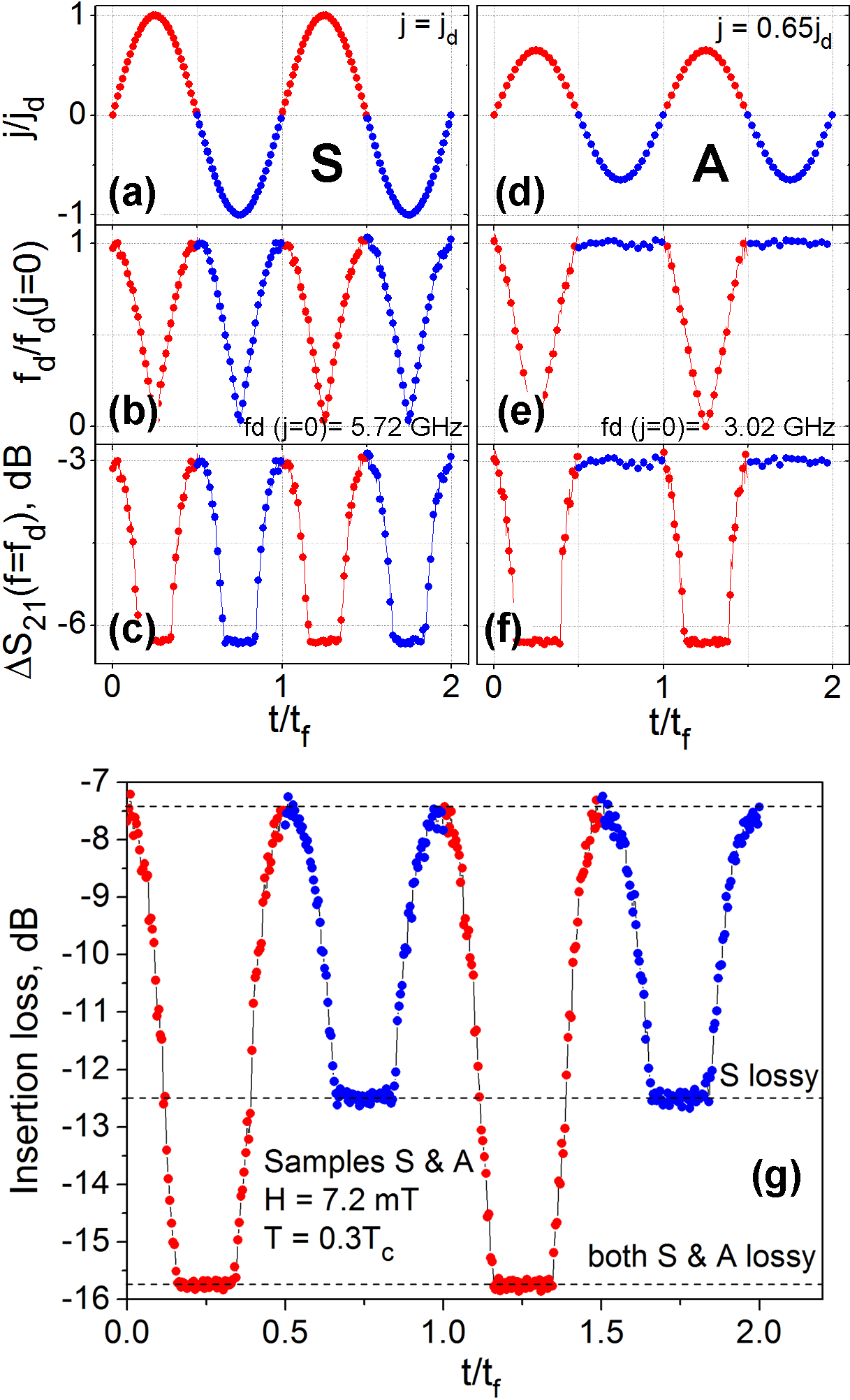}
    \caption{Microwave loss modulation in microstrips S (a-c) and A (d-f) at $H = 7.2$\,mT, $T = 0.3T_c$, and the excitation power $P = - 20$\,dBm. The adiabatic ac current ($3$\,Hz) in the close-to-critical (a) regime for sample S and in the critical regime for the weak slope of sample A leads  to the reduction of the depinning frequency [panels (b) and (e), respectively] and the modulation of the excess loss due to the vortices $\Delta S_{21}(f=f_d(j=0))$ in panels (c) and (f), respectively. The rectangular ``floors'' in panels (c) and (f) correspond to the maximal loss value $-6.4$\,dB in the flux-flow regime. In panel (g) is exemplified synthesis of three different insertion loss levels by a serial connection of samples S and A for an ac current $I = 50$\,mA and a frequency $3.02$\,GHz. After \cite{Dob15sbm}.}
    \label{fModul}
\end{figure}
A low-frequency ($3$\,Hz) ac current was used \cite{Dob15sbm} for tailoring the mw excess loss levels by vortices. The sine waves of the input quasistatic currents are shown in Fig. \ref{fModul}(a,d). The reduction of the depinning frequency upon ramping the current value is shown in Fig. \ref{fModul}(b,e). Panels (c) and (f) in Fig. \ref{fModul} present the time dependences of the excess loss $\Delta S_{21}(t, f=f_d)$ where the depinning frequency $f_d$ are chosen as \emph{carrier frequencies}. In particular, in Fig. \ref{fModul}(b) one sees that for sample S  in the critical-amplitude regime, the modulation of $f_d$ attains its maximal depth with a module-of-sine shape. Accordingly, $\Delta S_{21}$ is modulated for both ac halfwaves and attains the $-6.4$\,dB ``floors'' corresponding to the maximum excess loss level due to the vortex motion in the flux-flow regime. The modulation pattern for sample A in the critical current regime for the weak-slope WPP direction is substantially different. The depinning frequency is modulated during the positive ac halfwave (in red) only, whereas it remains constant during the negative ac halfperiod (in blue), see Fig. \ref{fModul}(e). In consequence of this, the excess loss $\Delta S_{21}(t)$ is observed during the positive ac halfwaves and is absent during the negative one. The shape of  $\Delta S_{21}(t)$  is nearly rectangular, that is a \emph{sine-to-rectangular pulse form conversion} takes place.

To exemplify the excess loss level synthesis, one considers the mw transmission through samples S and A connected in series. For definiteness,  the ac current $I$ with an amplitude of $50$\,mA is applied that, given the thicknesses of the samples, corresponds to the current density amplitude $0.48$\,MA/cm$^2$ for sample A and $0.83$\,MA/cm$^2$ for sample S. From the CVCs it follows that $j=0.48$\,MA/cm$^2 \simeq 0.52$\,MA/cm$^2=j_d^+$ for sample A and $j=0.83$\,MA/cm$^2 \simeq 0.75$\,MA/cm$^2=j_d$ for sample S.

The cumulative insertion loss in Fig. \ref{fModul}(g) is characterized by three different insertion loss levels, whereby the intermediate level is achieved in consequence of the combination of the lossy state of microstrip S and the low-lossy state of microstrip A. This can be understood as a result of summation of the individual excess loss levels in samples S and A with the help of panels (c) and (F) in Fig. \ref{fModul}, given the larger difference (about $5$\,dB) in $\Delta S_{21}$ for sample S at $3.02$\,GHz with respect to $-6.4$\,dB.

\section{Perspective}
To summarize, vortex dynamics in nanopatterned superconductors is a domain of intensive research. Among many interesting problems in this area a few ongoing challenging research lines should be emphasized. In particular, it was argued that the one-to-one mapping of a vortex to a particle breaks down at high velocities where vortex displacements take place at shorter time scales than the time to recover the superconducting order parameter along the vortex trajectory. The vortices then become elongated and turn into phase slip lines \cite{Lar75etp,Siv03prl,Sil10prl,Von11prl,Zol14ltp}. While it has been demonstrated that vortex guiding persists up to several GHz \cite{Wor12prb}, the crossover from the vortex-mediated flux transport to that via phase slips should be elaborated.

An interesting related phenomenon is the effect of micro\-wave-stimu\-lated superconductivity observed in both, flat \cite{Zol13ltp} and nanopatterned \cite{Lar15nsr} films. In addition, dynamic pinning landscapes, which can be provided by, e.\,g. a pulsed laser excitation of the superconductor surface, were predicted \cite{Zel15nsr} to support synchronization effects but not experimentally realized so far. New insights into vortex matter are also expected from studying it in more sophisticated nano-architectures \cite{Fom12nlt,Thu10nlt}.

It was reported that sub-terahertz transmission of a superconducting metamaterial can be modulated by passing electrical current through it \cite{Sav12prl}. Quantum metamaterials comprised of networks of superconducting qubits based on Josephson junctions are used for exploring quantum effects in meta-atoms \cite{Ust15met}. Tailoring microwave losses due to Abrikosov vortices in individual sections of superconducting transmission lines allows one to use them as building blocks for \emph{fluxonic metamaterials} \cite{Dob15sbm} enriching the large family of artificial media \cite{Zhe10sci}.

On the nanometa-materials science side a special interest lies in the use of focused electron beam induced deposition (FEBID) to directly write superconducting (S) nanostructures. At the time of writing, artificial FEBID-superconductors were limited to the systems MoC \cite{Mak14nsr} and Pb$_x$C$_y$O$_{\delta}$ \cite{Win14apl}, while further systems fabricated by FEBID are expected to be available in the years to come. The suitability of FEBID for the fabrication of ferromagnets (F) \cite{Lar14apl,Beg15nan,Dob15bjn} promises to bridge the gap to study spin-triplet proximity-induced superconductivity \cite{Buz05rmp,Ber05rmp} in S/F nanowire structures. Vortex matter in low-dimensional systems with proximity-induced superconductivity represents an intriguing \cite{Kop13etp}, but ``uncharted domain'' so far.

Finally, the description of the dynamics of Abrikosov vortices in superconductors with a washboard pinning landscape, relying upon the Langevin equation \cite{Cof12boo}, allows one to regard them as a ``playground'' to foresee the effects in related systems which require considering the dynamics of (quasi)particles in a periodic potential. Exemplary systems include Josephson junctions, superionic conductors \cite{Ful75prl}, ring laser gyroscopes~\cite{Cho85rmp}, the dynamics of spin \cite{Bar93prl} and charge density \cite{Zyb13prb} waves, phase-locking loops \cite{Ris89boo} in radioengineering, the motion of domain walls \cite{Per08prl}, magnetization dynamics \cite{Tit05prb} and the diffusion of colloidal particles in periodic structures \cite{Evs08pre}. For instance, molecular dynamics simulations \cite{Rei15prl} have recently been applied to spin textures termed \emph{skyrmions}  that were recently discovered in chiral magnets \cite{Muh09sci} and have particle-like properties and many similarities to superconducting vortices. In the presence of a washboard landscape, the force-velocity characteristic of the system has been predicted to exhibit Shapiro steps \cite{Rei15arx}, similar to the observations in superconducting vortex systems.

\clearpage
\section*{Acknowledgements}
The author is grateful to M. Huth for providing access to the instrumentation and discussing the experimental data. The author thanks V. A. Shklovskij for ideas and theoretical calculations that stimulated experiments reported in this work and for numerous fruitful discussions of their results. The author acknowledges the support of all colleagues who contributed at different stages of this work, namely R. Sachser for helping with nanopatterning and automating the data acquisition, E. Begun for depositing Co nanostripes, J. Franke for assembling the high-frequency sample probe and to M. Kompaniiets, M. Hanefeld, M. Z\"orb and L. K\"ohs for helping with selected measurements. This work was funded via DAAD scholarship A/08/96378, project NanoMag within the research collaboration NanoBiC, DFG projects DO1511/2-1 to 2-4, the Goethe University funding program ``Nachwuchswissenschaftler im Fokus'', Vereinigung von Freunden und F\"ordern der Goethe-Universit\"at and conducted within the framework of the COST Actions MP1201 (NanoSC-COST) and CM1301 (CELINA) of the European Cooperation in Science and Technology.
\vspace{3mm}

%\bibliographystyle{spbasic}      % basic style, author-year citations
%\bibliographystyle{spmpsci}      % mathematics and physical sciences
%\bibliographystyle{spphys}       % APS-like style for physics
%\bibliographystyle{elsarticle-num}
%\bibliography{D:/bibliobase/fluxonics,D:/bibliobase/proximity}

\section*{References}
\begin{thebibliography}{100}
\expandafter\ifx\csname url\endcsname\relax
  \def\url#1{\texttt{#1}}\fi
\expandafter\ifx\csname urlprefix\endcsname\relax\def\urlprefix{URL }\fi
\expandafter\ifx\csname href\endcsname\relax
  \def\href#1#2{#2} \def\path#1{#1}\fi

\bibitem{Shu37etf}
L.~V. Shubnikov, V.~I. Khotkevich, Y.~D. Shepelev, Y.~N. Ryabinin, Zh. Eksper.
  Teor. Fiz. 7 (1937) 221--237.

\bibitem{Shu08ujp}
L.~V. Shubnikov, V.~I. Khotkevich, Y.~D. Shepelev, Y.~N. Ryabinin, Ukr. J.
  Phys. 53 (2008) 42--52.

\bibitem{Abr57etp}
A.~A. Abrikosov, Sov. Phys. JETP. 5 (1957) 1174--1182.

\bibitem{Cam72aip}
A.~M. Campbell, J.~E. Evetts,
  \href{http://www.tandfonline.com/doi/abs/10.1080/00018737200101288}{Adv.
  Phys.} 21~(90) (1972) 199--428.

\bibitem{Bla94rmp}
G.~Blatter, M.~V. Feigel'man, V.~B. Geshkenbein, A.~I. Larkin, V.~M. Vinokur,
  \href{http://link.aps.org/doi/10.1103/RevModPhys.66.1125}{Rev. Mod. Phys.} 66
  (1994) 1125--1388.

\bibitem{Bra95rpp}
E.~H. Brandt, \href{http://stacks.iop.org/0034-4885/58/i=11/a=003}{Rep. Progr.
  Phys.} 58~(11) (1995) 1465--1594.

\bibitem{Nie69jap}
A.~K. Niessen, C.~H. Weijsenfeld,
  \href{http://link.aip.org/link/?JAP/40/384/1}{J. Appl. Phys.} 40~(1) (1969)
  384--393.

\bibitem{Mor70prl}
D.~D. Morrison, R.~M. Rose,
  \href{http://link.aps.org/doi/10.1103/PhysRevLett.25.356}{Phys. Rev. Lett.}
  25 (1970) 356--359.

\bibitem{Fio71prl}
A.~T. Fiory, \href{http://link.aps.org/doi/10.1103/PhysRevLett.27.501}{Phys.
  Rev. Lett.} 27 (1971) 501--503.

\bibitem{Fio78apl}
A.~T. Fiory, A.~F. Hebard, S.~Somekh,
  \href{http://link.aip.org/link/?APL/32/73/1}{Appl. Phys. Lett.} 32 (1978)
  73--75.

\bibitem{Mar75ssc}
P.~Martinoli, O.~Daldini, C.~Leemann, E.~Stocker, Solid State Commun. 17 (1975)
  205--209.

\bibitem{Git66prl}
J.~I. Gittleman, B.~Rosenblum,
  \href{http://link.aps.org/doi/10.1103/PhysRevLett.16.734}{Phys. Rev. Lett.}
  16 (1966) 734--736.

\bibitem{Mos10boo}
V.~V. Moshchalkov, R.~W\"ordenweber, M.~Lang, Nanoscience and
  Engineering in Superconductivity, Springer-Verlag, Berlin Heidelberg, 2010.

\bibitem{Luq07prb}
Q.~Lu, C.~J.~Olson Reichhardt, C.~Reichhardt,
  \href{http://link.aps.org/doi/10.1103/PhysRevB.75.054502}{Phys. Rev. B} 75
  (2007) 054502.

\bibitem{Kra78prl}
L.~Kramer, R.~J. Watts-Tobin,
  \href{http://link.aps.org/doi/10.1103/PhysRevLett.40.1041}{Phys. Rev. Lett.}
  40 (1978) 1041--1044.

\bibitem{Ros96prb}
E.~Rosseel, M.~Van~Bael, M.~Baert, R.~Jonckheere, V.~V. Moshchalkov,
  Y.~Bruynseraede,
  \href{http://link.aps.org/doi/10.1103/PhysRevB.53.R2983}{Phys. Rev. B} 53
  (1996) R2983--R2986.

\bibitem{Von11prl}
J.~Van~de Vondel, V.~N. Gladilin, A.~V. Silhanek, W.~Gillijns, J.~Tempere,
  J.~T. Devreese, V.~V. Moshchalkov,
  \href{http://link.aps.org/doi/10.1103/PhysRevLett.106.137003}{Phys. Rev.
  Lett.} 106 (2011) 137003.

\bibitem{Sil11sst}
A.\,Silva,\,N.\,Pompeo,\,S.\,Sarti,
  \href{http://stacks.iop.org/0953-2048/24/i=2/a=024018}{Supercond.\,Sci.\,Technol.}\,24\,(2011)\,024018.

\bibitem{Fom12nlt}
V. M. Fomin, R. Rezaev, O. Schmidt,
  \href{http://dx.doi.org/10.1021/nl203765f}{Nano Lett.} 12 (2012)
  1282--1287.

\bibitem{Cof12boo}
W.~T. Coffey, Y.~P. Kalmykov, The Langevin Equation, World Scientific, 2012.

\bibitem{Cof91prl}
M.~W. Coffey, J.~R. Clem,
  \href{http://link.aps.org/doi/10.1103/PhysRevLett.67.386}{Phys. Rev. Lett.}
  67 (1991) 386--389.

\bibitem{Pom08prb}
N.~Pompeo, E.~Silva,
  \href{http://link.aps.org/doi/10.1103/PhysRevB.78.094503}{Phys. Rev. B} 78
  (2008) 094503--1--10.

\bibitem{Maw97prb}
Y.~Mawatari, \href{http://link.aps.org/doi/10.1103/PhysRevB.56.3433}{Phys. Rev.
  B} 56 (1997) 3433--3437.

\bibitem{Maw99prb}
Y.~Mawatari, \href{http://link.aps.org/doi/10.1103/PhysRevB.59.12033}{Phys.
  Rev. B} 59 (1999) 12033--12038.

\bibitem{Shk99etp}
V.~A. Shklovskij, A.~K. Soroka, A.~A. Soroka, J. Exp. Theor. Phys. 89 (1999)
  1138--1153.

\bibitem{Shk02prb}
V.~A. Shklovskij, \href{10.1103/PhysRevB.65.092508}{Phys. Rev. B} 65 (2002)
  092508--1--4.

\bibitem{Shk06prb}
V.~A. Shklovskij, O.~V. Dobrovolskiy,
  \href{http://link.aps.org/doi/10.1103/PhysRevB.74.104511}{Phys. Rev. B} 74
  (2006) 104511.

\bibitem{Shk07pcs}
V.~A. Shklovskij, O.~V. Dobrovolskiy,
  \href{http://www.sciencedirect.com/science/article/pii/S0921453407006065}{Physica C} 460
  (2007) 1253.

\bibitem{Shk08prb}
V.~A. Shklovskij, O.~V. Dobrovolskiy,
  \href{http://link.aps.org/doi/10.1103/PhysRevB.78.104526}{Phys. Rev. B} 78
  (2008) 104526.

\bibitem{Shk09pcs}
V.~A. Shklovskij, O.~V. Dobrovolskiy,
  \href{http://stacks.iop.org/1742-6596/150/i=5/a=052241}{J. Phys.: Conf. Ser.}
  150~(5) (2009) 052241.

\bibitem{Shk09pcs2}
V.~A. Shklovskij,
  \href{http://dx.doi.org/10.1088/1742-6596/150/5/052240}{J. Phys.: Conf. Ser.}
  150~(5) (2009) 052240.

\bibitem{Shk09prb}
V.~A. Shklovskij, V.~V. Sosedkin,
  \href{http://link.aps.org/doi/10.1103/PhysRevB.80.214526}{Phys. Rev. B} 80
  (2009) 214526.

\bibitem{Shk09ltp}
V.~A. Shklovskij, D.~T.~B. Hop,
  \href{http://dx.doi.org/10.1063/1.3132742}{Low Temp. Phys.} 35
  (2009) 365.

\bibitem{Shk10ltp}
V.~A. Shklovskij, D.~T.~B. Hop,
  \href{http://dx.doi.org/10.1063/1.3292939}{Low Temp Phys.} 36
  (2010) 71.

\bibitem{Shk11prb}
V.~A. Shklovskij, O.~V. Dobrovolskiy,
  \href{http://link.aps.org/doi/10.1103/PhysRevB.84.054515}{Phys. Rev. B} 84
  (2011) 054515.

\bibitem{Dob12php}
O.~V. Dobrovolskiy, V.~A. Shklovskij, M.~Huth, Phys. Proc. 36 (2012) 7.

\bibitem{Shk12pcs}
V.~A. Shklovskij, O.~V. Dobrovolskiy,
  \href{http://stacks.iop.org/1742-6596/400/i=2/a=022108}{J. Phys.: Conf. Ser.}
  400~(2) (2012) 022108.

\bibitem{Shk12inb}
V.~A. Shklovskij, O.~V. Dobrovolskiy, Microwave Absorption by Vortices in
  Superconductors with a Washboard Pinning Potential, InTech, Rijeka, 2012,
  Ch.~11, pp. 263--288.

\bibitem{Shk13ltp}
V.~A. Shklovskij, O.~V. Dobrovolskiy,
  \href{http://link.aip.org/link/?LTP/39/120/1}{Low Temp. Phys.} 39 (2013)
  120.

\bibitem{Shk13snm}
V.~A. Shklovskij, O.~V. Dobrovolskiy, M.~Huth,
  \href{http://dx.doi.org/10.1007/s10948-012-1946-x}{J. Supercond. Nov.
  Magnet.} 26~(5) (2013) 2079--2083.

\bibitem{Shk14ltp}
V.~A. Shklovskij, J.-T. Seo,
  \href{http://dx.doi.org/10.1007/s10948-012-1946-x}{Low Temp. Phys.} 40 (2014) 1048.

\bibitem{Shk14pcm}
V.~A. Shklovskij, V.~V. Sosedkin, O.~V. Dobrovolskiy,
  \href{http://stacks.iop.org/0953-8984/26/i=2/a=025703}{J. Phys.: Cond. Matt.}
  26~(2) (2014) 025703.

\bibitem{Shk14phc}
V.~A. Shklovskij, O.~V. Dobrovolskiy,
  \href{http://www.sciencedirect.com/science/article/pii/S0921453414000501}{Physica
  C} 503 (2014) 128 -- 131.

\bibitem{Shk14pcs}
V.~A. Shklovskij, O.~V. Dobrovolskiy,
  \href{http://stacks.iop.org/1742-6596/507/i=1/a=012007}{J. Phys.: Conf. Ser.}
  507~(1) (2014) 012007.

\bibitem{Zhu03prb}
B.~Y. Zhu, F.~Marchesoni, V.~V. Moshchalkov, F.~Nori,
  \href{http://link.aps.org/doi/10.1103/PhysRevB.68.014514}{Phys. Rev. B} 68
  (2003) 014514--1--14.

\bibitem{Zhu04prl}
B.~Y. Zhu, F.~Marchesoni, F.~Nori,
  \href{http://link.aps.org/doi/10.1103/PhysRevLett.92.180602}{Phys. Rev.
  Lett.} 92 (2004) 180602.

\bibitem{Sor05phd}
O.~K. Soroka, Vortex dynamics in superconductors in the presence of anisotropic
  pinning, Ph.D. thesis, J. Gutenberg University (2005).

\bibitem{Wor06pcs}
R.~W\"ordenweber, J.~Sankarraj, P.~Dymashevski, E.~Hollmann,
  \href{http://www.sciencedirect.com/science/article/pii/S0921453405007987}{Physica
  C} 434~(1) (2006) 101--104.

\bibitem{Sil06inb}
E.~Silva, N.~Pompeo, S.~Sarti, C.~Amabile, Vortex State Microwave Response in
  Superconducting Cuprates, Nova Science, Hauppauge, NY, 2006, Ch.~1, pp.
  201--243.

\bibitem{Plo09tas}
B.~L.~T. Plourde,
  \href{http://ieeexplore.ieee.org/xpl/articleDetails.jsp?arnumber=5232856}{IEEE
  Trans. Appl. Supercond.} 19 (2009) 3698--3714.

\bibitem{Lee99nat}
C. Lee, B.~Janko, I.~Derenyi, A. Barabasi,
  \href{http://dx.doi.org/10.1038/22485}{Nature} 400 (1999) 337--340.

\bibitem{Han09rmp}
P.~H\"anggi, F.~Marchesoni,
  \href{http://link.aps.org/doi/10.1103/RevModPhys.81.387}{Rev. Mod. Phys.} 81
  (2009) 387--442.

\bibitem{Mos11boo}
V.~V. Moshchalkov, J.~Fritzsche, Nanostructured Superconductors, World
  Scientific, Singapore, 2011.

\bibitem{Jaq02apl}
D.~Jaque, E.~M. Gonzalez, J.~I. Martin, J.~V. Anguita, J.~L. Vicent,
  \href{http://link.aip.org/link/?APL/81/2851/1}{Appl. Phys. Lett.} 81 (2002)
  2851--2854.

\bibitem{Yuz99pcs}
Y.~Yuzhelevski, G.~Jung,
  \href{http://www.sciencedirect.com/science/article/pii/S0921453499001483}{Physica
  C} 314~(314) (1999) 163--171.

\bibitem{Sor07prb}
O.\,K.\,Soroka, V.\,A.\,Shklovskij, M.\,Huth,
  \href{http://link.aps.org/doi/10.1116/1.2955728}{Phys.\,Rev.\,B}\,76\,(2007)\,014504.

\bibitem{Wor12prb}
R.~W\"ordenweber, E.~Hollmann, J.~Schubert, R.~Kutzner, G.~Panaitov,
  \href{http://link.aps.org/doi/10.1103/PhysRevB.85.064503}{Phys. Rev. B} 85
  (2012) 064503.

\bibitem{Dob10sst}
O.~V. Dobrovolskiy, M.~Huth, V.~A. Shklovskij,
  \href{http://dx.doi.org/10.1088/0953-2048/23/12/125014}{Supercond. Sci.
  Technol.} 23~(12) (2010) 125014.

\bibitem{Dob12njp}
O.~V. Dobrovolskiy, E.~Begun, M.~Huth, V.~A. Shklovskij,
  \href{http://stacks.iop.org/1367-2630/14/i=11/a=113027}{New J. Phys.} 14~(11)
  (2012) 113027.

\bibitem{Ber97prl}
P.~Berghuis, E.~Di~Bartolomeo, G.~A. Wagner, J.~E. Evetts,
  \href{http://link.aps.org/doi/10.1103/PhysRevLett.79.2332}{Phys. Rev. Lett.}
  79 (1997) 2332--2335.

\bibitem{Cha98sst}
V.~V. Chabanenko, A.~A. Prodan, V.~A. Shklovskij, A.~V. Bondarenko, M.~A.
  Obolenskii, H.~Szymczak, S.~Piechota,
  \href{http://stacks.iop.org/0953-2048/11/i=10/a=052}{Supercond. Sci.
  Technol.} 11~(10) (1998) 1133--1136.

\bibitem{Pas99prl}
H.~Pastoriza, S.~Candia, G.~Nieva,
  \href{http://link.aps.org/doi/10.1103/PhysRevLett.83.1026}{Phys. Rev. Lett.}
  83 (1999) 1026--1029.

\bibitem{Dan00prb}
G.~D'Anna, V.~Berseth, L.~Forr\'o, A.~Erb, E.~Walker,
  \href{http://link.aps.org/doi/10.1103/PhysRevB.61.4215}{Phys. Rev. B} 61
  (2000) 4215--4221.

\bibitem{Yuz99prb}
Y.~Yuzhelevski, G.~Jung, C.~Camerlingo, M.~Russo, M.~Ghinovker, B.~Y. Shapiro,
  \href{http://link.aps.org/doi/10.1103/PhysRevB.60.9726}{Phys. Rev. B} 60
  (1999) 9726--9733.

\bibitem{Hut02afm}
M.~Huth, K.~Ritley, J.~Oster, H.~Dosch, H.~Adrian,
  \href{http://dx.doi.org/10.1002/1616-3028(20020517)12:5<333::AID-ADFM333>3.0.CO;2-C}{Adv.
  Func. Mat.} 12~(5) (2002) 333--338.

\bibitem{Sil11sut}
A.~V. Silhanek, V.~N. Gladilin, V.~de~Vondel.~J., B.~Raes, G.~W. Ataklti,
  W.~Gillijns, J.~Tempere, J.~T. Devreese, V.~V. Moshchalkov,
  \href{http://stacks.iop.org/0953-2048/24/i=2/a=024007}{Supercond. Sci.
  Technol.} 24~(2) (2011) 024007.

\bibitem{Utk08vst}
I.~Utke, P.~Hoffmann, J.~Melngailis,
  \href{http://dx.doi.org/10.1116/1.2955728}{J. Vac. Sci. B} 26 (2008)
  1197--1276.

\bibitem{Hut12bjn}
M.~Huth, F.~Porrati, C.~Schwalb, M.~Winhold, R.~Sachser, M.~Dukic, J.~Adams,
  G.~Fantner, \href{http://dx.doi.org/10.3762/bjnano.3.70}{Beilstein J.
  Nanotechnol.} 3 (2012) 597--619.

\bibitem{Dob11snm}
O.~V. Dobrovolskiy, M.~Huth, V.~A. Shklovskij,
  \href{http://dx.doi.org/10.1007/s10948-010-1055-7}{J. Supercond. Nov.
  Magnet.} 24 (2011) 375--380.

\bibitem{Nak93jpd}
Y.~Nakayama, T.~Makabe, \href{http://stacks.iop.org/0022-3727/26/i=10/a=034}{J.
  Phys. D: Appl. Phys.} 26~(10) (1993) 1769.

\bibitem{Pau04prb}
A.~Pautrat, J.~Scola, C.~Goupil, C.~Simon, C.~Villard, B.~Domeng\`es, Y.~Simon,
  C.~Guilpin, L.~M\'echin,
  \href{http://link.aps.org/doi/10.1103/PhysRevB.69.224504}{Phys. Rev. B} 69
  (2004) 224504--1--5.

\bibitem{Lia05vst}
T.~Liang, E.~Frendberg, B.~Lieberman, A.~Stivers,
  \href{http://scitation.aip.org/content/avs/journal/jvstb/23/6/10.1116/1.2062428}{J.
  Vac. Sci. Technol. B} 23~(6) (2005) 3101--3105.

\bibitem{Kom14jap}
M.~Kompaniiets, O.~V. Dobrovolskiy, C.~Neetzel, E.~Begun, F.~Porrati,
  W.~Ensinger, M.~Huth,
  \href{http://scitation.aip.org/content/aip/journal/jap/116/7/10.1063/1.4893549}{J.
  Appl. Phys.} 116~(7) (2014) 073906--1--10.

\bibitem{Dan06lan}
C.~Danelon, C.~Santschi, J.~Brugger, H.~Vogel,
  \href{http://pubs.acs.org/doi/abs/10.1021/la061321c}{Langmuir} 22~(25) (2006)
  10711--10715.

\bibitem{Gab10nan}
M.~Gabureac, L.~Bernau, I.~Utke, G.~Boero,
  \href{http://dx.doi.org/10.1088/0957-4484/21/11/115503}{Nanotechnology}
  21~(11) (2010) 115503.

\bibitem{Sch10sen}
C.~H. Schwalb, C.~Grimm, M.~Baranowski, R.~Sachser, F.~Porrati, H.~Reith,
  P.~Das, J.~M\"uller, F.~V\"olklein, A.~Kaya, M.~Huth,
  \href{http://dx.doi.org/10.3390/s101109847}{Sensors} 10~(11) (2010)
  9847--9856.

\bibitem{Mak14nsr}
K.~Makise, K.~Mitsuishi, M.~Shimojo, B.~Shinozaki,
  \href{http://dx.doi.org/10.1038/srep05740}{Sci. Rep.} 4 (2014) 5740.

\bibitem{Win14apl}
M.~Winhold, P.~M. Weirich, C.~H. Schwalb, M.~Huth,
  \href{http://scitation.aip.org/content/aip/journal/apl/105/16/10.1063/1.4898819}{Appl.
  Phys. Lett.} 105~(16) (2014) 162603.

\bibitem{Lar14apl}
A.~Lara, O.~V. Dobrovolskiy, J.~L. Prieto, M.~Huth, F.~G. Aliev,
  \href{http://scitation.aip.org/content/aip/journal/apl/105/18/10.1063/1.4900789}{Appl.
  Phys. Lett.} 105~(18) (2014) 182402.

\bibitem{Kei06nat}
R.~S. Keizer, S.~T.~B. Goennenwein, T.~M. Klapwijk, G.~Miao, G.~Xiao, A.~Gupta,
  \href{http://dx.doi.org/10.1038/nature04499}{Nature} 439~(7078) (2006)
  825--827.

\bibitem{All05sci}
D.~A. Allwood, \href{http://dx.doi.org/10.1126/science.1108813}{Science}
  309~(5741) (2005) 1688--1692.

\bibitem{Par08sci}
S.~S.~P. Parkin, M.~Hayashi, L.~Thomas,
  \href{http://dx.doi.org/10.1126/science.1145799}{Science} 320 (2008)
  190--194.

\bibitem{Bel12rsi}
L.~M. Belova, O.~Hellwig, E.~Dobisz, E.~Dan~Dahlberg,
  \href{http://scitation.aip.org/content/aip/journal/rsi/83/9/10.1063/1.4752225}{Rev.
  Sci. Instr.} 83~(9) (2012) 093711--1--4.

\bibitem{Tak06jap}
Y.~K. Takahashi, K.~Hono, S.~Okamoto, O.~Kitakami,
  \href{http://scitation.aip.org/content/aip/journal/jap/100/7/10.1063/1.2355442}{J.
  Appl. Phys.} 100~(7) (2006) 074305--1--7.

\bibitem{Bri11prl}
L.~O'Brien, D.~Petit, E.~R. Lewis, R.~P. Cowburn, D.~E. Read, J.~Sampaio, H.~T.
  Zeng, A.-V. Jausovec,
  \href{http://link.aps.org/doi/10.1103/PhysRevLett.106.087204}{Phys. Rev.
  Lett.} 106 (2011) 087204.

\bibitem{Vel08mmm}
M.~Velez, J.~I. Martin, J.~E. Villegas, A.~Hoffmann, E.~M. Gonzalez, J.~L.
  Vicent, I.~K. Schuller,
  \href{http://www.sciencedirect.com/science/article/pii/S0304885308007014}{J.
  Magn. Magnet. Mat.} 320 (2008) 2547--2562.


\bibitem{Buz05rmp}
A.~Buzdin, \href{http://dx.doi.org/10.1103/RevModPhys.77.935}{Rev. Mod. Phys.}
  77~(3) (2005) 935--976.

\bibitem{Ber05rmp}
F.\,Bergeret,\,A.\,Volkov,\,K.\,Efetov,
  \href{http://dx.doi.org/10.1103/RevModPhys.77.1321}{Rev.\,Mod.\,Phys.}\,77\,(2005)\,1321--1373.

\bibitem{Wan10nat}
J.\,Wang, M.\,Singh, M.\,Tian, N.\,Kumar, B. Liu, C. Shi, J.\,K. Jain, N.\,Sam\-arth,
  T.~E. Mallouk, M.~H.~W. Chan,
  \href{http://dx.doi.org/10.1038/nphys1621}{Nature Phys.} 6~(5) (2010)
  389--394.

\bibitem{Kom14apl}
M.~Kompaniiets, O.~V. Dobrovolskiy, C.~Neetzel, F.~Porrati, J.~Br\"otz,
  W.~Ensinger, M.~Huth, \href{http://dx.doi.org/10.1063/1.4863980}{Appl. Phys.
  Lett.} 104 (2014) 052603.

\bibitem{Dob11pcs}
O.~V. Dobrovolskiy, E.~Begun, M.~Huth, V.~A. Shklovskij, M.~I. Tsindlekht,
  \href{http://www.sciencedirect.com/science/article/pii/S0921453411003327}{Physica
  C} 471~(15-16) (2011) 449--452.

\bibitem{Pla15prc}
H. Plank,
  \emph{Private communication} (2015).

\bibitem{Beg15nan}
E.~Begun, O.~V. Dobrovolskiy, M.~Kompaniiets, C.~Gspan, H.~Plank, M.~Huth,
  \href{http://stacks.iop.org/0957-4484/26/i=7/a=075301}{Nanotechnology} 26~(7)
  (2015) 075301.

\bibitem{Dob15bjn}
O.~V. Dobrovolskiy, M.~Kompaniiets, R.~Sachser, F.~Porrati, C.~Gspan, H.~Plank,
  M.~Huth, \href{http://dx.doi.org/10.3762/bjnano.6.109}{Beilstein J.
  Nanotech.} 6 (2015) 1082--1090.

\bibitem{Dob12ppa}
O.~V. Dobrovolskiy, M.~Huth, V.~A. Shklovskij, Acta Phys. Pol. A 121 (2012)
  82--84.

\bibitem{Sil08apl}
A.~V. Silhanek, J.~V. de~Vondel, V.~V. Moshchalkov, A.~Leo, V.~Metlushko,
  B.~Ilic, V.~R. Misko, F.~M. Peeters,
  \href{http://link.aip.org/link/?APL/92/176101/1}{Appl. Phys. Lett.} 92~(17)
  (2008) 176101--1--2.

\bibitem{Dob13pcs}
O.~V. Dobrovolskiy, E.~Begun, M.~Huth, V.~A. Shklovskij,
  \href{http://www.sciencedirect.com/science/article/pii/S0921453413001123}{Physica
  C} 494~(0) (2013) 102 -- 105.

\bibitem{Dob12tsf}
O.~V. Dobrovolskiy, M.~Huth,
  \href{http://www.sciencedirect.com/science/article/pii/S0040609012005718}{Thin
  Solid Films} 520 (2012) 5985--5990.

\bibitem{Dob15mst}
O.~V. Dobrovolskiy, J.~Franke, M.~Huth,
  \href{http://stacks.iop.org/0957-0233/26/i=3/a=035502}{Meas. Sci. Technol.}
  26~(3) (2015) 035502.

\bibitem{Lar75etp}
A.~I. Larkin, Y.~N. Ovchinnikov,
  \href{http://www.jetp.ac.ru/cgi-bin/index/e/41/5/p960?a=list}{J. Exp. Theor.
  Phys.} 41 (1975) 960--965.

\bibitem{Dob15snm}
O.~V. Dobrovolskiy., J. Supercond. Nov. Magnet. 28 (2015) 469--473.

\bibitem{Sol14prb}
P.-d.-J. Cuadra-Solis, A.~Garcia-Santiago, J.~M. Hernandez, J.~Tejada,
  J.~Vanacken, V.~V. Moshchalkov,
  \href{http://link.aps.org/doi/10.1103/PhysRevB.89.054517}{Phys. Rev. B} 89
  (2014) 054517.

\bibitem{Lar15nsr}
A.~Lara, F.~G. Aliev, A.~V. Silhanek, V.~V. Moshchalkov,
  \href{http://dx.doi.org/10.1038/srep09187}{Sci. Rep.} 5 (2015) 9187.

\bibitem{Dob15apl}
O.~V. Dobrovolskiy, M.~Huth,
  \href{http://scitation.aip.org/content/aip/journal/apl/106/14/10.1063/1.4917229}{Appl.
  Phys. Lett.} 106 (2015) 142601.

\bibitem{Jan06prb}
D.~Janju\ifmmode \check{s}\else \v{s}\fi{}evi\ifmmode~\acute{c}\else
  \'{c}\fi{}, M.~S. Grbi\ifmmode~\acute{c}\else \'{c}\fi{},
  M.~Po\ifmmode~\check{z}\else \v{z}\fi{}ek, A.~Dul\ifmmode \check{c}\else
  \v{c}\fi{}i\ifmmode~\acute{c}\else \'{c}\fi{}, D.~Paar, B.~Nebendahl,
  T.~Wagner, \href{http://link.aps.org/doi/10.1103/PhysRevB.74.104501}{Phys.
  Rev. B} 74 (2006) 104501--1--7.


\bibitem{Pom13snm}
N.~Pompeo, K.~Torokhtii, C.~Meneghini, S.~Mobilio, R.~Loria, C.~Cirillo,
  E.~Ilyina, C.~Attanasio, S.~Sarti, E.~Silva,
  \href{http://dx.doi.org/10.1007/s10948-012-2057-4}{J. Supercond. Nov.
  Magnet.} 26~(5) (2013) 1939--1943.

\bibitem{Poz11boo}
D.~M. Pozar, Microwave engineering, J. Wiley \& Sons, Inc., NY, 2011.

\bibitem{Zai03prb}
A.~G. Zaitsev, R.~Schneider, G.~Linker, F.~Ratzel, R.~Smithey, J.~Geerk,
  \href{http://link.aps.org/doi/10.1103/PhysRevB.68.104502}{Phys. Rev. B} 68
  (2003) 104502.

\bibitem{Dob15sbm}
O.\,V. Dobrovolskiy, M.~Huth, V.\,A. Shklovskij,
  \href{http://dx.doi.org/10.1063/1.4934487}{Appl. Phys. Lett.} 107 (2015) 162603.

\bibitem{Siv03prl}
A.~G. Sivakov, A.~M. Glukhov, A.~N. Omelyanchouk, Y.~Koval, P.~M\"uller, A.~V.
  Ustinov, \href{http://link.aps.org/doi/10.1103/PhysRevLett.91.267001}{Phys.
  Rev. Lett.} 91 (2003) 267001--1--4.

\bibitem{Sil10prl}
A.~V. Silhanek, M.~V. Milo\ifmmode \check{s}\else
  \v{s}\fi{}evi\ifmmode~\acute{c}\else \'{c}\fi{}, R.~B.~G. Kramer, G.~R.
  Berdiyorov, J.~Van~de Vondel, R.~F. Luccas, T.~Puig, F.~M. Peeters, V.~V.
  Moshchalkov,
  \href{http://link.aps.org/doi/10.1103/PhysRevLett.104.017001}{Phys. Rev.
  Lett.} 104 (2010) 017001.

\bibitem{Zol14ltp}
I.~V. Zolochevskii,
  \href{http://scitation.aip.org/content/aip/journal/ltp/40/10/10.1063/1.4900695}{Low
  Temp. Phys.} 40~(10) (2014) 867--892.

\bibitem{Zol13ltp}
I.~V. Zolochevskii, \href{http://dx.doi.org/10.1063/1.4813655}{Low Temp. Phys.} 39 (2013) 571.

\bibitem{Zel15nsr}
Z. Jelic, M.~Milosevic, A.~V. Silhanek, J.~Van~de Vondel, \href{http://dx.doi.org/10.1038/srep14604}{Sci. Rep.} 5 (2015) 14604.


\bibitem{Thu10nlt}
D. J. Thurmer, C. C. B. Bufon, Ch. Deneke, O. G. Schmidt,
\href{http://dx.doi.org/10.1021/nl1022145}{Nano Lett.} 10 (2010) 3704-3709.

\bibitem{Sav12prl}
V.~Savinov, V.~A. Fedotov, S.~M. Anlage, P.~A.~J. de~Groot, N.~I. Zheludev,
  \href{http://link.aps.org/doi/10.1103/PhysRevLett.109.243904}{Phys. Rev.
  Lett.} 109 (2012) 243904.

\bibitem{Ust15met}
A.~V. Ustinov, in: Proc. 9th Internat. Congr. Adv.
  Electromagn. Mater. Microw. Optics, Oxford, UK, 2015.

\bibitem{Zhe10sci}
N.~I. Zheludev,
  \href{http://www.sciencemag.org/content/328/5978/582.short}{Science}
  328~(5978) (2010) 582--583.

\bibitem{Kop13etp}
N.~B. Kopnin, I.~M. Khaymovich, A.~S. Mel'nikov, J. Exp. Theor. Phys. 117
  (2013) 418--438.

\bibitem{Ful75prl}
P.~Fulde, L.~Pietronero, W.~R. Schneider, S.~Str\"assler,
  \href{http://link.aps.org/doi/10.1103/PhysRevLett.35.1776}{Phys. Rev. Lett.}
  35 (1975) 1776--1779.

\bibitem{Cho85rmp}
W.~W. Chow, J.~Gea-Banacloche, L.~M. Pedrotti, V.~E. Sanders, W.~Schleich,
  M.~O. Scully, \href{http://link.aps.org/doi/10.1103/RevModPhys.57.61}{Rev.
  Mod. Phys.} 57 (1985) 61--104.

\bibitem{Bar93prl}
E.\,Barthel, G.\,Kriza, G.\,Quirion, P.\,Wzietek, D.\,J\'erome, J.\,B.\,Christen\-sen,
  M.~J\o{}rgensen, K.~Bechgaard,
  \href{http://link.aps.org/doi/10.1103/PhysRevLett.71.2825}{Phys. Rev. Lett.}
  71 (1993) 2825--2828.

\bibitem{Zyb13prb}
S.~G. Zybtsev, V.~Y. Pokrovskii,
  \href{http://link.aps.org/doi/10.1103/PhysRevB.88.125144}{Phys. Rev. B} 88
  (2013) 125144.

\bibitem{Ris89boo}
H.~Risken, {The Fokker-Planck Equation}, Springer, Berlin, 1989.

\bibitem{Per08prl}
A.~P\'erez-Junquera, V.~I. Marconi, A.~B. Kolton, L.~M. \'Alvarez-Prado,
  Y.~Souche, A.~Alija, M.~V\'elez, J.~V. Anguita, J.~M. Alameda, J.~I.
  Mart\'in, J.~M.~R. Parrondo,
  \href{http://link.aps.org/doi/10.1103/PhysRevLett.100.037203}{Phys. Rev.
  Lett.} 100 (2008) 037203.

\bibitem{Tit05prb}
S.~V. Titov, H.~Kachkachi, Y.~P. Kalmykov, W.~T. Coffey,
  \href{http://link.aps.org/doi/10.1103/PhysRevB.72.134425}{Phys. Rev. B} 72
  (2005) 134425.

\bibitem{Evs08pre}
M.~Evstigneev, O.~Zvyagolskaya, S.~Bleil, R.~Eichhorn, C.~Bechinger,
  P.~Reimann, \href{http://link.aps.org/doi/10.1103/PhysRevE.77.041107}{Phys.
  Rev. E} 77 (2008) 041107.

\bibitem{Rei15prl}
C.~Reichhardt, D.~Ray, C.~J.~O. Reichhardt,
  \href{http://link.aps.org/doi/10.1103/PhysRevLett.114.217202}{Phys. Rev.
  Lett.} 114 (2015) 217202.

\bibitem{Muh09sci}
S.~M\"uhlbauer, B.~Binz, F.~Jonietz, C.~Pfleiderer, A.~Rosch, A.~Neubauer,
  R.~Georgii, P.~B\"oni,
  \href{http://www.sciencemag.org/content/323/5916/915.abstract}{Science}
  323~(5916) (2009) 915--919.

\bibitem{Rei15arx}
C.~Reichhardt, C.~J. Olson~Reichhardt, \href{http://arxiv.org/abs/1507.03023}{arXiv:1507.03023v1} (2015).

\end{thebibliography}

\end{document}